\documentclass[twocolumn,aps,amsmath,amssymb]{revtex4-1}
\usepackage{graphicx,bm,amsmath,amssymb,natbib,color}
\usepackage[normalem]{ulem}
\usepackage{cancel}
\usepackage[dvipsnames]{xcolor}
\begin{document}

\title{Circuit-QED with phase-biased Josephson weak links}
\author{C. Metzger\textsuperscript{1}, Sunghun Park\textsuperscript{2}, L.~Tosi\textsuperscript{1,3}}
\author{C. Janvier\textsuperscript{1}}
\altaffiliation[Present address: ]{MUQUANS, Institut d'Optique d'Aquitaine, rue Fran\c{c}ois Mitterrand, 33400, Talence, France}
\author{A.~A.~Reynoso\textsuperscript{3}, M.~F.~Goffman\textsuperscript{1}, C. Urbina\textsuperscript{1}, A. Levy Yeyati\textsuperscript{2}, and H. Pothier\textsuperscript{1}}
\email[Corresponding author~: ]{hugues.pothier@cea.fr}
\affiliation{\textsuperscript{1}Quantronics group, Service de Physique de l'\'Etat Condens\'e (CNRS,
UMR\ 3680), IRAMIS, CEA-Saclay, Universit\'e Paris-Saclay, 91191 Gif-sur-Yvette, France\\
\textsuperscript{2}Departamento de F\'{\i}sica Te\'orica de la Materia Condensada, Condensed Matter Physics Center (IFIMAC) and Instituto Nicol\'as Cabrera, Universidad Aut\'onoma de Madrid, 28049 Madrid, Spain\\
\textsuperscript{3}Centro At\'omico Bariloche and Instituto Balseiro, CNEA, CONICET, 8400 San Carlos de Bariloche, R\'io Negro, Argentina\\}
\date{\today}

\begin{abstract}
By coupling a superconducting weak link to a microwave resonator, recent experiments probed the spectrum and achieved the quantum manipulation of Andreev states in various systems. However, the quantitative understanding of the response of the resonator to changes in the occupancy of the Andreev levels, which are of fermionic nature, is missing. Here, using Bogoliubov-de Gennes formalism to describe the weak link and a general formulation of the coupling to the resonator,
we calculate the shift of the resonator frequency as a function of the levels occupancy and describe how transitions are induced by phase or electric field microwave drives. 
We apply this formalism to analyze recent experimental results obtained using circuit-QED techniques on superconducting atomic contacts and semiconducting nanowire Josephson junctions. 
\end{abstract}
\maketitle

\section{Introduction}
Circuit-QED (cQED) was born from the transcription of the concepts of Cavity Quantum Electrodynamics \cite{Raimond2001,Miller2005}, which describes the strong coupling of atoms to the photons of a cavity, to superconducting circuits in which the discrete spectrum of collective electromagnetic modes mimics the energy levels of an atom \cite{Blais2004,Wallraff2004}. These techniques are nowadays at the core of quantum information processing with superconducting circuits \cite{Wendin2017,Blais2020}. Furthermore, the coupling of superconducting cavities with a variety of systems is now intensively used to probe and manipulate other degrees of freedom at the single quantum level \cite{Bruhat2016,Cottet2017,Clerk2020}. These can be either collective modes, like the modes of a nanomechanical oscillator \cite{Aspelmeyer2014}, or microscopic ones like the spin of an electron in semiconducting quantum dots or in defects in an insulator \cite{Burkard2019}. 
Most superconducting circuits probed so far using cQED correspond to the former case, as the relevant modes are surface plasmons in nonlinear oscillators, rendered dissipationless by the existence of the superconducting gap. A very interesting exception arises in phased-biased Josephson weak links or in quantum dots with superconducting electrodes, where localized discrete subgap states (Andreev bound states (ABS) \cite{Kulik1970,Beenakker1991a,Furusaki1991,Bagwell1992} or Yu-Shiba-Rusinov states \cite{Martin-Rodero2011,Kirsanskas2015}) develop, giving rise to a truly atomic-like spectrum of levels for fermionic quasiparticles. 
Two types of transitions, both conserving fermion parity, can be driven in an Andreev system: pair transitions, in which two quasiparticles are excited at once, and single quasiparticle transitions that correspond to atomic-like transitions of a quasiparticle from one Andreev level to another. 
Pair transitions, first observed in atomic contacts \cite{Bretheau2013}, give rise to the Andreev qubit \cite{Desposito2001,Zazunov2003,Janvier2015}. Single quasiparticle transitions, first observed in InAs weak links \cite{Tosi2019}, offer a route, alternative to quantum dots, to couple a single fermionic spin with a microwave resonator and develop an Andreev spin qubit \cite{Chtchelkatchev2003,Padurariu2010,Park2017, Hays2019, Hays2020}. Most of these experiments were performed using cQED techniques  \cite{Janvier2015,Hays2017,Tosi2019,Hays2019, Hays2020}.
The aim of the present work is to develop a theory of the coupling of a microwave resonator to a multi-level many-body fermionic system of Andreev levels, able to describe the main features of the spectra measured in those recent experiments, in particular the intensity of the transition lines, and analyze the possible existence of selection rules associated to the spin.

The article is organised as follows. In Sec.~\ref{nodrive} we introduce the theoretical model and discuss the resonator frequency shift in the absence of driving fields. This shift is given, up to a prefactor, by the imaginary part of the weak link admittance, a quantity that has been computed for zero-length weak links in Refs.~\cite{Kos2013,Peng2016}, using linear response theory. Here, we start from the microscopic Bogoliubov-de Gennes (BdG) equations for a weak link of arbitrary length, and express the shift in the resonator frequency as a function of the occupancy of individual Andreev levels. In Sec.~\ref{gate-flux-drive} we introduce a model to describe the driving through either an ac flux or an ac gate voltage and find under which conditions spin-non-conserving transitions can occur. In Sec.~\ref{driven-shift} we derive the resonator frequency shifts in the presence of driving fields. In Sec.~\ref{comparison-AC}, we compare the predictions of the theory to experimental data on superconducting atomic contacts and nanowire weak links.

\begin{figure}[t]
\includegraphics[width=0.9\columnwidth]{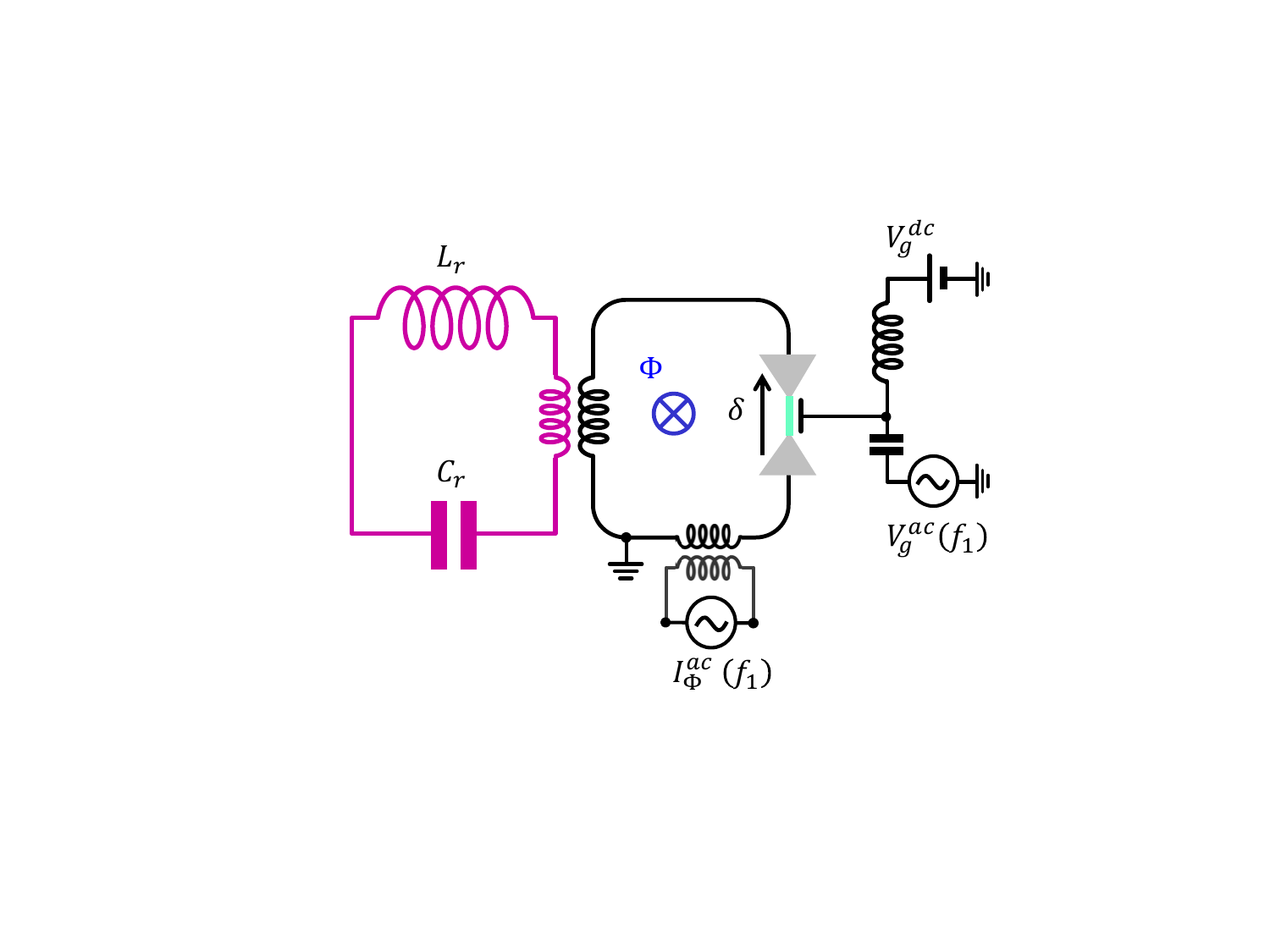}
\caption{Schematics of a superconducting weak link (light green) placed in a superconducting loop and inductively coupled to a  microwave resonator, represented as a lumped elements LC circuit. The superconducting phase $\delta$ across the weak link is imposed by the magnetic flux $\Phi$ threading the loop. 
The resonator frequency depends on the occupancy of the Andreev states in the weak link. Transitions between Andreev states can be driven by an ac signal either through a gate ($V_{g}^{ac}$) or a flux line ($I_{\Phi}^{ac}$). }
\label{Fig:Fig1}
\end{figure}

\begin{figure*}[t!]
\includegraphics[width=1.8\columnwidth]{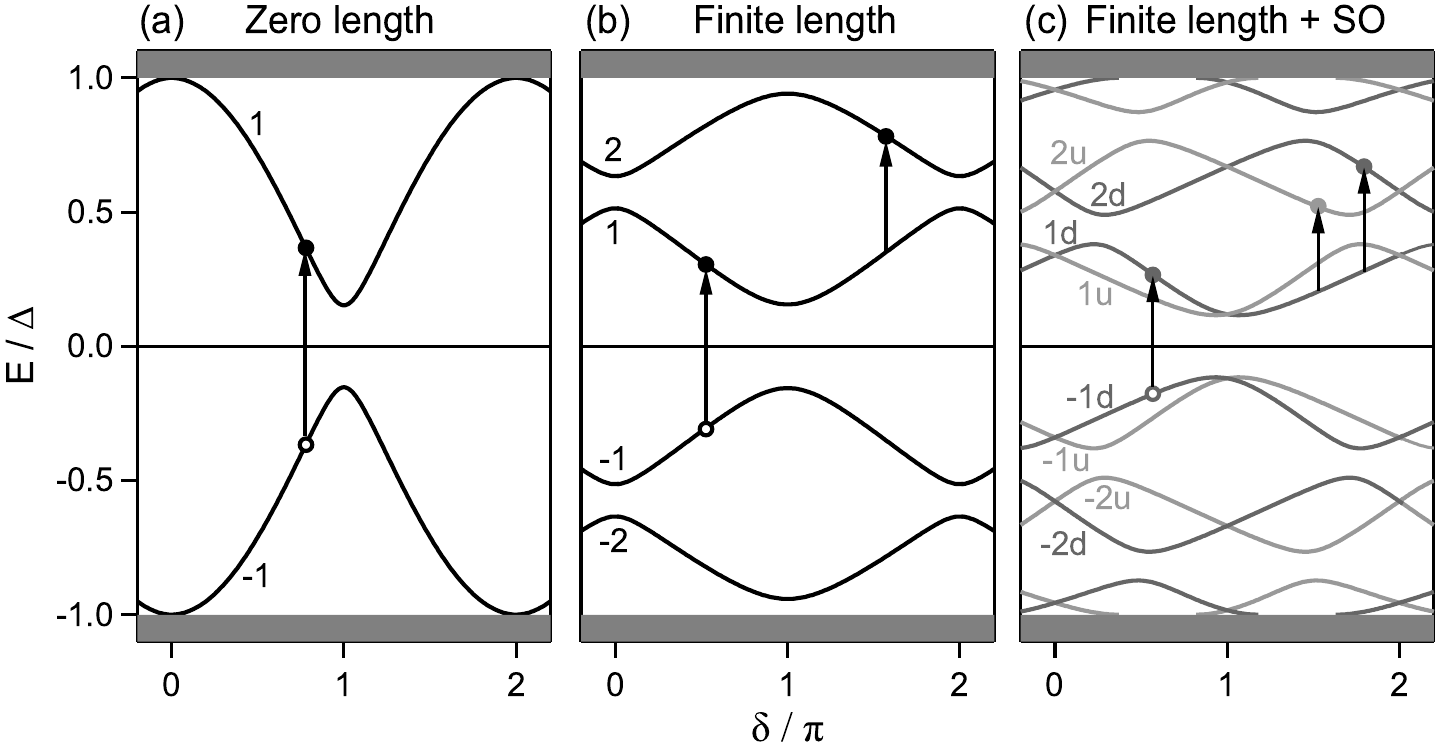}
\caption{Phase dependence of Andreev states, as obtained by solving the Bogoliubov-de Gennes equations, for  
(a) a zero-length weak link, 
(b) a finite-length weak link, and 
(c) a finite-length weak link in presence of spin-orbit coupling. In (a,b), all lines are spin-degenerate. In (c), dark and light gray lines correspond to Andreev levels with different pseudospins. 
In the ground state, all levels with negative energy are occupied. 
Two types of transitions can occur: pair transitions, leading to two additional excitations, are represented with arrows crossing the Fermi energy (in (a), leftmost arrow in (b) and in (c)). Single particle transitions are possible when quasiparticle excitations are already present in the system. They correspond to atomic-like transitions between two levels both at either positive or negative energies (rightmost arrow in (b), rightmost arrows in (c)).}
\label{Fig:Fig2}
\end{figure*}

\section{Resonator-weak link coupling} 
\label{nodrive}

The system we consider comprises a weak link of length $L$ embedded in a superconducting loop that is inductively coupled to a microwave resonator (Fig.~\ref{Fig:Fig1}). The microwave cavity is represented as a lumped-element LC resonator with bare resonance frequency $f_r=(2\pi\sqrt{L_rC_r})^{-1}$. Introducing the photon annihilation (creation) operators $a$ ($a^{\dagger}$), it can be described by the Hamiltonian $\hat{H}_{r}=h f_{r} a^{\dagger} a.$ 

The dc phase difference $\delta=2\pi\Phi/\Phi_0$ across the weak link, with $\Phi_0=h/2e$ the flux quantum, is imposed by the magnetic flux $\Phi$ threading the superconducting loop (we assume in this section that the loop inductance is negligible). An electrostatic gate fixes with its dc value $V_g$ the electrochemical potential in the weak link. We consider applying a microwave drive either on the gate voltage $V_g^{\text{ac}}$ or on the flux $\Phi^{\text{ac}}$.

The Hamiltonian for the weak link can be written in the form
\begin{equation}
\hat{H}_0 (\delta) = \frac{1}{2} \int dx \; \hat{\Psi}^{\dagger}(x) {\cal H}_0 (\delta) \hat{\Psi}(x), \label{WLH}
\end{equation}
where $\hat{\Psi}(x) = \left(\psi_{\uparrow}(x),\psi_{\downarrow}(x),\psi^{\dagger }_{\downarrow}(x),-\psi^{\dagger}_{\uparrow}(x)\right)^T$ is the Nambu bispinor field operator and $x$ is the position along the weak link. We denote by $|\Phi_{i\sigma}\rangle$ the eigenstates of the Bogoliubov-de Gennes (BdG) equation ${\cal H}_0(\delta) |\Phi_{i\sigma}\rangle = E_{i\sigma} |\Phi_{i\sigma}\rangle$, which correspond to Andreev states when $|E_{i\sigma}|<\Delta$, where $\Delta$ is the superconducting gap in the leads. In this notation the subscript $i\sigma$ refers to the level $i$ with spin $\sigma$. Levels labeled with positive $i$ are above the Fermi level. Due to the  electron-hole symmetry implicit in the BdG formalism, each state $i\sigma$ is associated to a state with opposite spin at opposite energy $-i \bar{\sigma}.$ Notice that when spin-orbit interaction is at play, as can be the case in a nanowire weak link, spin is no longer a good quantum number and $\sigma$ has to be understood as a pseudospin index noted ``u'' or ``d'' in the following. Otherwise $\sigma$ corresponds to spin up ($\uparrow$) or spin down ($\downarrow$). In Fig.~\ref{Fig:Fig2} we show the energy of the Andreev states arising in a 
weak link with a single occupied channel,
for three cases of increasing complexity. (a) For a zero-length junction, there is only one pair of subgap spin-degenerate levels $i=\pm1$. Each level can be occupied by 0, 1 or 2 quasiparticles. This case describes well atomic contacts weak links with length $L \ll \xi$, where $\xi$ is the superconducting coherence length. For ballistic conduction channels, $\xi=\hbar v_F /\Delta$, where $v_F$ is the Fermi velocity at the weak link. (b) For finite length weak links, without spin-orbit coupling, the parameter $\lambda=L/ \xi$ determines how many spin-degenerate Andreev pairs appear in the gap: depending on $\delta$ and channel transmission $\tau$, this number is $1+\lfloor 2\lambda / \pi \rfloor$ or $2+\lfloor 2\lambda / \pi \rfloor$ ($\lfloor x \rfloor$ is the integer part of $x$). For the parameters of Fig.~\ref{Fig:Fig2}b ($\lambda=1.7,\tau=0.97$), $i=\pm1,\pm2$. Compared to short junctions, a new type of excitation arises: a quasiparticle in state $1$ can absorb a photon and be excited to state $2$. 
(c) The spin character of these excitations becomes relevant when spin-orbit interaction is present in the weak link, which leads to a lifting of the levels spin-degeneracy  when $\delta \neq 0, \pi$. This lifting results from a spin-dependent Fermi velocity, leading to different values of $\lambda$ for the two spin textures \cite{Park2017,Tosi2019}. In Fig.~\ref{Fig:Fig2}(c), the gray level of the lines encodes the state pseudospin. Among all possible transitions, some conserve the pseudospin, others do not. This regime describes well InAs nanowire weak links \cite{Tosi2019,Hays2019}.

The coupling between resonator and weak link occurs through current fluctuations in the resonator, assumed to be in its ground state, which induce phase fluctuations across the weak link, so that $\delta \rightarrow \delta + \hat{\delta}_r$, where $\hat{\delta}_r = \delta_{\text{zp}} (a+a^{\dagger})$ with $\delta_{\text{zp}}$ the amplitude of zero-point phase fluctuations. In accordance with experiments, we assume 
$\delta_{\text{zp}} \ll 1$. In Ref.~\cite{Park2020}, by expanding the Hamiltonian $\hat{H}_0$ up to second order in $\hat{\delta}_r$, we derived a general expression to compute the photon-number-dependent shift of the energy of an Andreev level and, more importantly here, the frequency shift of the resonator when a single level $|\Phi_{i\sigma}\rangle$ is occupied (we write here explicitly the spin indices) :

\begin{widetext}
\begin{equation}
\frac{h \delta f_r^{(i\sigma)}}{\delta^2_{\text{zp}}} =  E''_{i\sigma} + \sum_{j\sigma' \neq i\sigma} {\cal M}_{i\sigma,j\sigma'}^2 \left(\frac{2}{E_{i\sigma ,j\sigma'}}-\frac{1}{E_{i\sigma ,j\sigma'} -  h f_r} - \frac{1}{E_{i\sigma, j\sigma'} + h f_r} \right)= E''_{i\sigma}+ \sum_{j\sigma' \neq i\sigma}{\mathcal{V}}_{i\sigma, j\sigma'},
\label{fShift-Single}
\end{equation}
\end{widetext}
where we have introduced the transition energies $E_{i\sigma,j\sigma'}=E_{j\sigma'}-E_{i\sigma},$ the curvature $E''_{i\sigma}=\partial^2 E_{i\sigma}/\partial \delta^2$ and ${\cal M}_{i\sigma,j\sigma'}=|\langle \Phi_{i}| {\cal H}'_0|\Phi_{j}\rangle|$ the modulus of the matrix element of the current operator ${\cal H}'_0=\partial {\cal H}_0 /\partial \delta$ between states $i\sigma$ and $j\sigma'.$ 
The coupling strength $g_{i\sigma,j\sigma'}$ is related to ${\cal M}_{i\sigma,j\sigma'}$ by $\hbar g_{i\sigma,j\sigma'}=\delta_{\text{zp}} {\cal M}_{i\sigma,j\sigma'}.$ The term associated to virtual transitions from $i \sigma$ to $j \sigma'$ is noted $\mathcal{V}_{i\sigma, j\sigma'}.$ We note that the terms 
\begin{equation*}
 E''_{i\sigma} + \sum_{j\sigma' \neq i\sigma} {\cal M}_{i\sigma,j\sigma'}^2 \left(\frac{2}{E_{i\sigma ,j\sigma'}}\right),
\end{equation*}
in Eq.~(\ref{fShift-Single}) arise from the ${\cal H}''_0=\partial^2 {\cal H}_0/\partial \delta^2$ term in the expansion of Hamiltonian (\ref{WLH}), whereas the remaining terms
\begin{equation*}
\sum_{j\sigma' \neq i\sigma} {\cal M}_{i\sigma,j\sigma'}^2 \left(-\frac{1}{E_{i\sigma ,j\sigma'} -  h f_r} - \frac{1}{E_{i\sigma, j\sigma'} + h f_r} \right),
\end{equation*}
that result from the ${\cal H}'_0$ term can also be derived from the Jaynes-Cummings Hamiltonian \cite{Park2020}.
On the right-hand-side of Eq.~(\ref{fShift-Single}), one obtains, with a prefactor $\varphi_0^2 \omega_r,$ the imaginary part of the contribution of state $i \sigma$ to the weak link admittance (here, $\omega_r=2\pi f_r$). It  has two terms: one from the zero frequency response $1/L_J^{(i \sigma)}\omega_r$, with $L_J^{(i \sigma)}=\varphi_0^2/E''_{i\sigma}$  the Josephson inductance of state $i \sigma$ ($\varphi_0 = \Phi_0/2\pi$); the  second one from the response at finite frequency \cite{Kos2013}.

According to Ref.~\cite{Park2020} two different regimes can be identified: the {\it adiabatic} one, which occurs when $h f_r \ll |E_{i\sigma,j\sigma'}|$ for all $i\sigma,j\sigma'$, leading to $h\delta f_r^{(i\sigma)} \propto E''_{i\sigma}$; and the {\it dispersive} one when $h f_r \sim |E_{i\sigma,j\sigma'}|$ for a set of $i\sigma,j\sigma'$, in which case the terms which involve exchange of virtual photons dominate. We shall analyze the occurrence of the two regimes for the examples presented below.

It can be seen from Eq.~(\ref{fShift-Single}) that all transitions which couple a given state $i\sigma$ with other states $j\sigma'$ via ${\cal H}'_0$ are relevant to calculate the shift of the resonator. Note that $j\sigma'$ cannot be taken equal to $-i\bar{\sigma}$ since electron-hole symmetry anticommutes with ${\cal H}'_0$. 

\textbf{Fermionic statistics:} The actual resonator frequency shift is determined not by a single but by all Andreev levels which are populated in a given many-body state of the weak link. We consider first the ground state $|g\rangle$, in which all negative energy levels are occupied. The frequency shift is then 
\begin{equation}
\delta f_r^{|g\rangle} = \frac{1}{2}\sum_{i< 0,\sigma } \delta f_r^{(i\sigma)} \;,\label{fShift-Ground}
\end{equation}
where the factor $1/2$ compensates for the redundancy of the BdG description. Note how we differentiate in the notation the shift $\delta f_r^{(i\sigma)}$ (with parentheses) associated to the occupancy of a single level $i\sigma $ and the shift $\delta f_r^{|\Psi\rangle}$ (with a ket) associated to a many-body state $|\Psi\rangle$.
When combining Eq.~\eqref{fShift-Single} and Eq.~\eqref{fShift-Ground}, and taking into account that ${\mathcal{V}}_{i\sigma, j\sigma'}=-{\mathcal{V}}_{j\sigma', i\sigma}$, only virtual transitions to positive energy levels contribute
\begin{equation}
\frac{h \delta f_r^{|g\rangle}}{\delta_{\text{zp}}^2} =  E''_{|g\rangle} + \frac{1}{2}\sum_{\substack{i < 0, \sigma \\j  > 0, \sigma'}} \mathcal{V}_{i \sigma, j \sigma'},\label{fShift-Ground2}
\end{equation}
where $E_{|g\rangle}=(1/2)\sum_{i< 0,\sigma}E_{i\sigma}$ is the energy of the ground state. Further simplification occurs in the absence of a magnetic field and in the presence of a mirror symmetry, where the operator ${\cal H}'_0$ does not mix opposite pseudospins ($\sigma$ and $\bar{\sigma}$) \cite{Park2017}, so that $\mathcal{V}_{i \sigma, j \bar{\sigma}}=0.$ 

Any many-body state $|\Psi\rangle$ can be built by creating electron-like $\gamma^{\dagger}_{i\sigma}|g\rangle$ ($i>0$) or hole-like $\gamma^{}_{i\sigma}|g\rangle$ ($i<0$) excitations from the ground state. Here $\gamma^{\dagger}_{i\sigma}$ is the Bogoliubov quasiparticle creation operator. Notice also that $\gamma^{\dagger}_{-i\bar{\sigma}} = -s \gamma_{i\sigma}$ due to double counting in the semiconducting picture that we are using, where $s=1(-1)$ for $\sigma=u(d)$. The frequency shift in $|\Psi\rangle$ is
\begin{equation}
\delta f_r^{|\Psi\rangle} =  \delta f_r^{|g\rangle}+ \sum_{i>0,\sigma} \left[ n_{i\sigma} \delta f_r^{(i\sigma)} - \left(1-n_{-i\sigma}\right) \delta f_r^{(-i\sigma)}\right]\;,\label{fShift-MB}
\end{equation}
where $n_{i\sigma}=0,1$ is the occupancy of the state $i\sigma$. More generally,  $n_{i\sigma}$ has to be understood as the average occupancy of this state. 
The number of fermionic quasiparticle excitations in the weak link given by $N_{|\Psi\rangle}=\sum_{i>0,\sigma} \left[ n_{i\sigma} +1 - n_{-i\sigma} \right]$ can be even or odd, but states with different parity are not coupled by photons. 

We now illustrate these ideas in a simple case.

\begin{figure}[t]
\includegraphics[width=1\columnwidth]{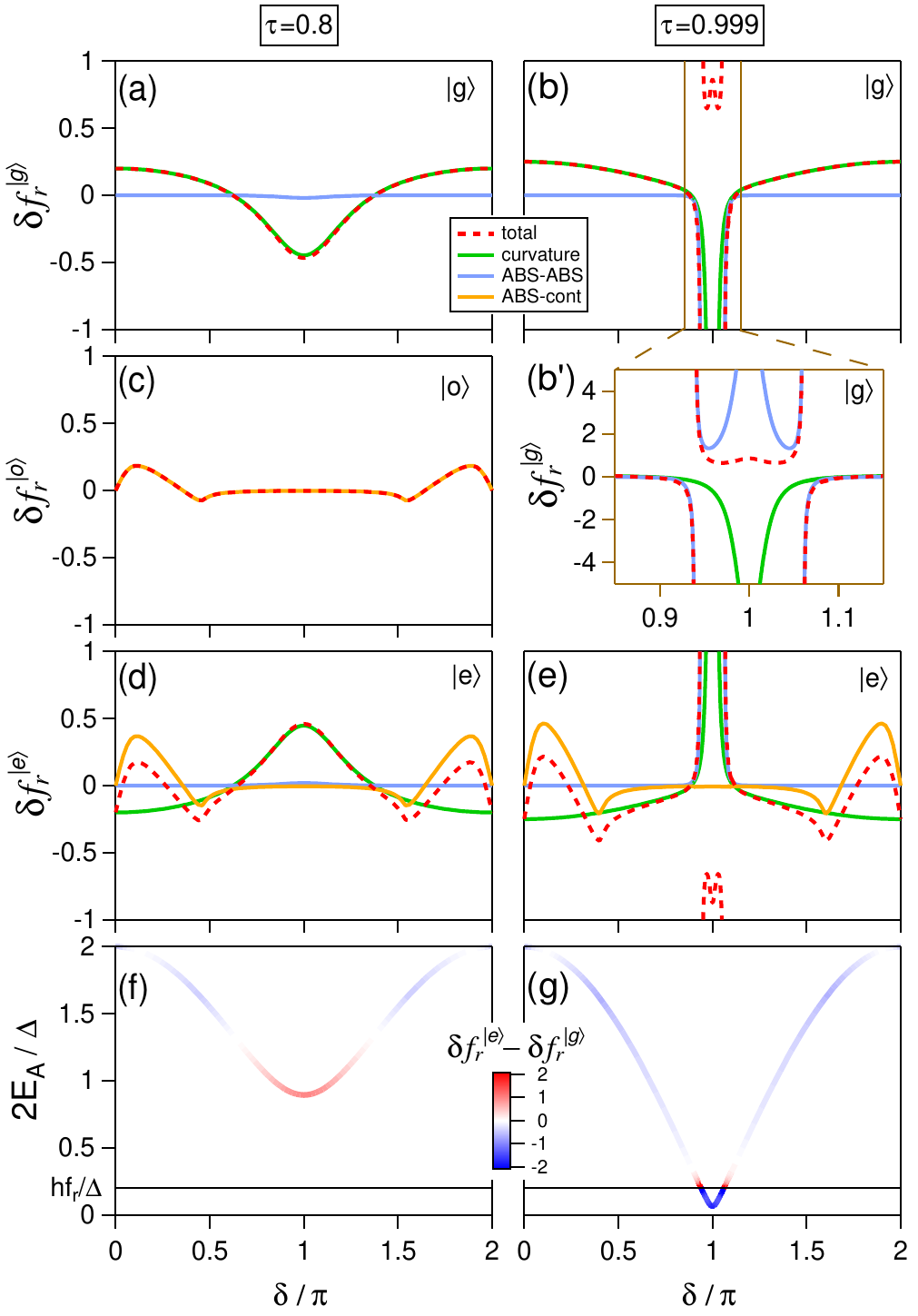}
\caption{
Zero-length one-channel junction. (a-e) Resonator frequency shifts (in units of $\delta_{\text zp}^2\Delta/h$) for two values of the channel transmission $\tau$, and (f,g) transition energy $2E_A$ with color-coded frequency change when driving the system from $|g\rangle$ to $|e\rangle$, all as a function of the phase $\delta$. The bare resonator frequency, shown as a black line in (f,g), was taken at $f_r=0.2\Delta/h.$ In the left panels, $\tau=0.8$, and the transition frequency is always larger than $h f_r$; in the right ones, $\tau=0.999$, and $2E_A$ crosses $h f_r$. Total frequency shifts $\delta f_r^{|g\rangle,|o\rangle,|e\rangle}$ in state $|g\rangle$ (a,b,b'), $|o\rangle$ (c) and $|e\rangle$ (d,e) are shown with dashed red lines. They are decomposed into three contributions: states' curvature $E"_{|g\rangle,|o\rangle,|e\rangle}$ (green lines), virtual transitions among Andreev levels (blue lines) and virtual transitions from Andreev levels to continuum levels (orange lines). In each panel, the contributions that have a negligible contribution are not shown, for clarity.}
\label{Fig:ZeroLength}
\end{figure}

\textbf{Zero-length junctions:} In the limit $L\rightarrow0$, accessed experimentally in superconducting atomic contacts, the BdG equation can be solved analytically. For a single conduction channel of transmission $\tau$, the resulting pair of Andreev states within the gap, represented in Fig.~\ref{Fig:Fig2}(a), has energies $E_{\pm 1,\sigma}=\pm E_{A}(\delta)=\pm\Delta \sqrt{1-\tau \sin^2(\delta/2)}$ \cite{Beenakker1991a, Furusaki1991, Bagwell1992}. In Fig.~\ref{Fig:ZeroLength} the frequency shift of a resonator at $f_r=0.2\Delta/h$ as given by Eq.~(\ref{fShift-MB}) is shown for $\tau=0.8$ and $\tau=0.999$, and for three many-body states: 
the ground state $|g\rangle,$ the odd parity state $|o\rangle$  obtained by creation of one quasiparticle $|o\rangle=|1\sigma\rangle =\gamma^{\dagger}_{1\sigma}|g\rangle,$ and
the lowest-in-energy excited state with even parity  $|e\rangle=|1\mathord{\uparrow}1\mathord{\downarrow}\rangle =\gamma^{\dagger}_{1\uparrow}\gamma^{\dagger}_{1\downarrow}|g\rangle$. 
In each state, the resonator frequency shift (dashed red line in Fig.~\ref{Fig:ZeroLength}) results from the sum of four contributions. The first one corresponds to the contribution of the curvature $E''_{|\Psi\rangle}$ of the many-body state $|\Psi\rangle$ (green lines in Fig.~\ref{Fig:ZeroLength}). The second one (blue lines) is associated with virtual transitions between the Andreev levels -1 and +1 (see arrow in Fig.~\ref{Fig:Fig2}(a)), coupled by the matrix element of ${\cal H}'_0$ \cite{Zazunov2014,Janvier2015}
\begin{equation}
{\cal M}={\cal M}_{-1\sigma,1\sigma}=  \frac{\Delta\sqrt{1-\tau}}{2} \left(\frac{\Delta}{E_A}-\frac{E_A}{\Delta}\right).
\label{AC-matelem}
\end{equation}
The corresponding term reads
\begin{equation}
\mathcal{V}_{-1\sigma,1\sigma}=\frac{{\cal M}^2}{h}\left(\frac{2}{f_A}-\frac{1}{f_A-f_r}-\frac{1}{f_A+f_r}  \right),
\label{nu-AC}
\end{equation}
with $f_A=2E_A/h$ the Andreev transition frequency.

The third type of contribution (orange lines), is associated with virtual transitions between an Andreev level and states in the continuum $\cal{C}^{-}$ at energies $E<-\Delta$ or $\cal{C}^{+}$ at energies $E>\Delta$. Using the expressions for the matrix elements of ${\cal H}'_0$ given in Refs.~\cite{Olivares2014,Zazunov2014}, and introducing a broadening of $10^{-3} \Delta$, one finds that the associated shift grows positive from $\delta=0$, presents a maximum, and exhibits a negative dip when $\Delta-E_A=h f_r$. This is characteristic of a threshold behavior associated with the continuum, also discussed in Ref.~\cite{Kos2013}.

The last contribution, which results from virtual transitions from states in $\cal{C}^{-}$ to states in $\cal{C}^{+}$, is negligible.

In the ground state (Fig.~\ref{Fig:ZeroLength}(a) and (b)), the level -1 and all levels of $\cal{C}^{-}$ are doubly occupied, so that the factor $\frac{1}{2}$ in Eq.~(\ref{fShift-Ground2}) cancels out with a factor 2 for the spin, 
and 
\begin{equation}
\frac{h \delta f_r^{|g\rangle}}{\delta_{\text{zp}}^2} =  E''_{|g\rangle} + \sum_{\substack{i < 0\\j  > 0}} \mathcal{V}_{i, j},
\end{equation}
where we dropped the spin indices since ${\cal H}'_0$ conserves the spin  for zero-length junctions. In the zero-length limit, the energy of the states in the continuum does not depend on phase \cite{Levchenko2006}, and $E''_{|g\rangle}=-E''_A.$ The second term reads
\begin{equation}
\sum_{\substack{i < 0\\j  > 0}} \mathcal{V}_{i, j}= \mathcal{V}_{-1,1}+\sum_{i\in\cal{C}^{-}}\mathcal{V}_{i,1}+\sum_{j\in\cal{C}^{+}}\mathcal{V}_{-1,j}+\sum_{\substack{i \in\cal{C}^{-}\\j\in\cal{C}^{+}}} \mathcal{V}_{i, j}.
\label{Eq:secondterm}
\end{equation}
Since $\mathcal{V}_{i, j}=-\mathcal{V}_{j, i}$ and $\mathcal{V}_{i, j}=-\mathcal{V}_{-i, -j},$ one obtains, neglecting virtual transitions from $\cal{C}^{-}$ to $\cal{C}^{+}$ (last term in Eq.~(\ref{Eq:secondterm})), 
\begin{equation}
\frac{h\delta f_r^{|g\rangle}}{\delta_{\text{zp}}^2} \approx - E''_{A} +\mathcal{V}_{-1,1}+2\sum_{j\in\cal{C}^{+}}\mathcal{V}_{-1,j}
.\label{AC-fShift-Ground2}
\end{equation}
In practice, because of the large energy  $E_{-1,j}>E_A+\Delta$ for a transition to $\cal{C}^{+}$, the last term in Eq.~\ref{AC-fShift-Ground2} can always be neglected
\begin{equation}
\frac{h\delta f_r^{|g\rangle}}{\delta_{\text{zp}}^2} \approx - E''_{A} + \frac{{\cal M}^2}{h}\left(\frac{2}{f_A}-\frac{1}{f_A-f_r}-\frac{1}{f_A+f_r} \right).
\end{equation}
When $f_A\gg f_r$, the three terms from $\mathcal{V}_{-1,1}$ compensate and the frequency shift is entirely due to  $E''_{A}$, as shown in Fig.~\ref{Fig:ZeroLength}(a) and far from $\delta=\pi$ in Fig.~\ref{Fig:ZeroLength}(b). 
When $|f_r-f_A| \ll \Delta$ there is a compensation between $-E''_{A}$ and ${\cal M}^2\frac{2}{h f_A}$ (green and blue lines in Fig.~\ref{Fig:ZeroLength}(b,b')), \textit{i.e.} the contribution due to $\mathcal{H}''_0$ vanishes \cite{Park2020}, and the frequency shift is essentially the one that can be derived from the Jaynes-Cummings Hamiltonian \cite{Johansson2006,Zueco2009}
\begin{equation}
\delta f_r^{|g\rangle,\text{JC}} =-\left(\frac{g(\delta)}{2\pi}\right)^2 \left(\frac{1}{f_A-f_r}+\frac{1}{f_A+f_r}  \right),
\label{JC}
\end{equation}
with $g(\delta)=\cal{M} \delta_{\text {zp}}/\hbar.$ At the scale of Fig.~\ref{Fig:ZeroLength}(b'), $\delta f_r^{|g\rangle,\text{JC}}$ and the exact $\delta f_r^{|g\rangle}$ coincide within the linewidth. The rotating-wave approximation (RWA), which consists in neglecting the second term:
\begin{equation}
\delta f_r^{|g\rangle,\text{RWA}} = - \frac{\left(g(\delta)/2\pi\right)^2}{f_{A} - f_r},
\label{RWA}
\end{equation}
overestimates the little bump of $\delta f_r^{|g\rangle}$ at $\delta=\pi$ by a factor $\sim 2.$

The odd parity state $|o\rangle=\gamma^{\dagger}_{1\sigma}|g\rangle$ has energy $E_g + E_{A}=0$. The shift of the resonator in this case is
\begin{equation}
\delta f_r^{|o\rangle}=\delta f_r^{|g\rangle}+\delta f_r^{(1)}.
\end{equation}
Using 
\begin{equation}
\frac{h\delta f_r^{(1)}}{\delta_{\text{zp}}^2} =  E''_A+\mathcal{V}_{1, -1}
+\sum_{j  \in \mathcal{C}^{-}} \mathcal{V}_{1, j}
+\sum_{j \in \mathcal{C}^{+}} \mathcal{V}_{1, j},
\end{equation}
and $\mathcal{V}_{1, j}=-\mathcal{V}_{-1, j},$ one obtains 
\begin{equation}
\frac{}{}h\delta f_r^{|o\rangle} \approx \sum_{j\in\cal{C}^{+}}\left(\mathcal{V}_{-1,j}+\mathcal{V}_{1,j}\right)
.\label{AC-fShift-odd}
\end{equation}
Although this state does not disperse with $\delta$ ($E_{|o\rangle}'=0$), one obtains a finite shift associated to transitions from Andreev states to continuum states, which becomes sizable when $E_A$ approaches $\Delta$ (see Fig.~\ref{Fig:ZeroLength}(c)).

The excited state $|e\rangle=\gamma^{\dagger}_{1\uparrow}\gamma^{\dagger}_{1\downarrow}|g\rangle$ has energy $E_g +2E_{1\sigma}=E_{A}$. The shift in this state (shown in  Fig.~\ref{Fig:ZeroLength}(d,e)) is
\begin{equation}
\delta f_r^{|e\rangle}=\delta f_r^{|g\rangle}+2\delta f_r^{(1)},
\end{equation}
and one gets
\begin{equation}
\frac{h\delta f_r^{|e\rangle}}{\delta_{\text{zp}}^2} \approx   E''_{A} -\mathcal{V}_{-1,1}+2\sum_{j\in\cal{C}^{+}}\mathcal{V}_{1,j}
.\label{AC-fShift-excited}
\end{equation}
When $E_A \ll \Delta$ the continuum contributions can be neglected and $\delta f_r^{|e\rangle} \approx -\delta f_r^{|g\rangle}$. This is no longer the case when $E_A$ approaches $\Delta,$ a situation in which both  $2\sum_{j\in\cal{C}^{+}}\mathcal{V}_{1,j}$ and $ E''_{A}$ contribute to the shift, as shown in Fig.~\ref{Fig:ZeroLength}(d,e).

In spectroscopy experiments \cite{Janvier2015}, transitions $|g\rangle \rightarrow |e\rangle$ are observed. The frequency shift that governs the measured signal is $\Delta f_r=\delta p_{|e\rangle} \left(  \delta f_r^{|e\rangle} -\delta f_r^{|g\rangle} \right)$, with $\delta p_{|e\rangle}$ the population change in the excited state due to the microwave excitation. In Fig.~\ref{Fig:ZeroLength}(f,g), $\delta f_r^{|e\rangle} -\delta f_r^{|g\rangle}$ is encoded in the color of the line showing $2E_A(\delta)$. For $\tau=0.8,$ it is dominated by the curvature term $2E''_A$, except when $E_A$ approaches $\Delta$ and virtual transitions enter in $\delta f_r^{|e\rangle}$. For $\tau=0.999$, close to $\delta=\pi$, the terms associated to virtual transitions between ABS causes a change of sign of $\Delta f_r$ when $2E_A$ crosses $hf_r,$ as expressed by the dispersive approximation (Eq.~\eqref{JC}).

These results coincide with those obtained from a linear response derivation of the admittance of the zero-length weak link in Ref.~\cite{Kos2013}, using the relation \begin{equation}
\frac{h \delta f_r^{|\Psi\rangle}}{\delta^2_{\text{zp}}} =\varphi_0^2 \omega_r \mathrm{Im}\left( \frac{i}{L_J^{|\Psi\rangle} \omega_r}+ Y^{|\Psi\rangle}(\omega_r)\right),
\label{fShift-linearResponse}
\end{equation}
where $L_J^{|\Psi\rangle}=\varphi_0^2 / E''_{|\Psi\rangle}$ and $Y^{|\Psi\rangle}$ are the Josephson inductance and the finite frequency admittance of the weak link in state $|\Psi\rangle$.

\textbf{Finite-length junctions, with spin-resolved levels:}
\begin{figure*}[t]
\includegraphics[width=2.05\columnwidth]{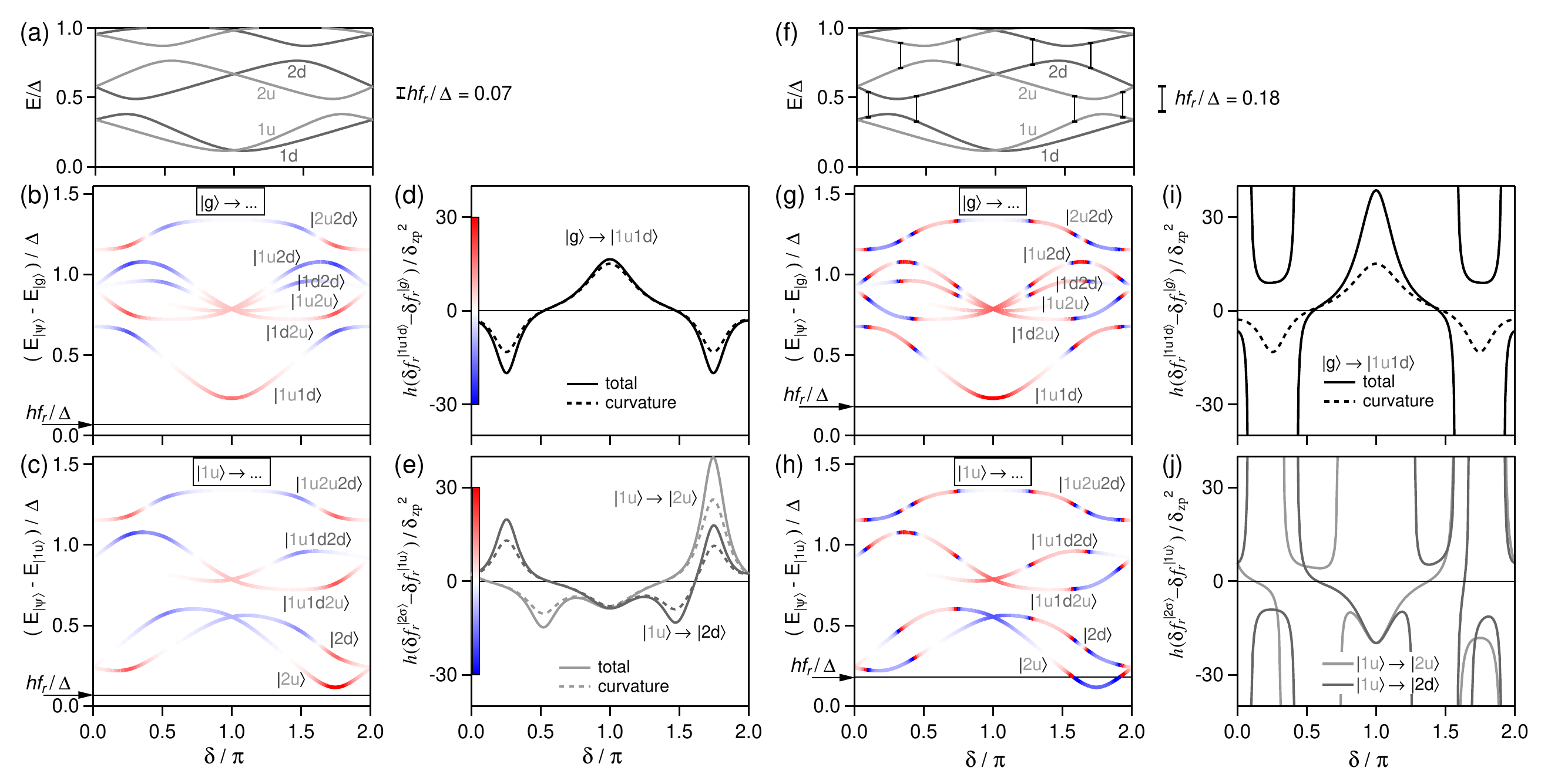}
\caption{Resonator frequency shift for finite-length junction with spin-orbit coupling (same parameters as in Fig.~\ref{Fig:Fig2}(c)). 
(a-e) Resonator at $f_r=0.07 \Delta/h$; (f-j) $f_r=0.18 \Delta/h$. (a,f) Spectrum of Andreev states at positive energies, with bars indicating places where energy difference between levels of same spin is equal to $hf_r$. (b,g) Energy of transitions from state $|g\rangle$; (c,h) \textit{idem} from state $|1u\rangle$. The color of the lines encodes the resonator frequency difference between initial and final state (colorscale on y-axis of (d) and (e)). (d,e) Resonator frequency shift difference between $|g\rangle$ and $|1u1d\rangle$ (pair transition), and contribution of the states' curvature (dashed line). (e,j) \textit{Idem} for the transitions from $|1u\rangle$ to $|2u\rangle$ or $|2d\rangle$ (single particle transitions). The resonator shifts are calculated from matrix  elements obtained with a tight-binding model.}
\label{Fig:FigLongJunction}
\end{figure*}
The situation becomes richer when there are several Andreev levels within the gap as in Fig.~\ref{Fig:Fig2}(b,c). Furthermore, in the presence of spin-orbit (Fig.~\ref{Fig:Fig2}(c)) the subgap states are spin split, which gives rise to a larger number of possible transitions.
The terms $\mathcal{V}_{-i\sigma, j\sigma}$ depend on the matrix elements of $\mathcal{H}'_0$ which do not have analytic forms in this case. They can be obtained by solving numerically the BdG equation, for which we use two complementary approaches: the scattering model of Ref.~\cite{Tosi2019} and a discretized tight-binding model of the nanowire (see Appendices \ref{app:scattering} and \ref{app:TB} for technical details). As these methods rely on different approximations one cannot expect a one-to-one correspondence of their results. For instance, the scattering method is based on a linearization of the electrons and holes dispersion relations around the Fermi level (Andreev approximation) which is not assumed in the tight-binding model. On the other hand the tight-binding model only includes two sites to describe the nanowire cross-section. We have checked, however, that the methods yield qualitatively similar results for the limits where their approximations are both valid.

In Fig.~\ref{Fig:FigLongJunction}, we show the predictions for the frequency shifts in the case of a weak link with three spin-split manifolds of Andreev levels (same parameters as Fig.~\ref{Fig:Fig2}(c)), at zero Zeeman field. Two values of the bare resonator frequency are considered: $f_r=0.07 \Delta/h$ (panels (a--e)) and $f_r=0.18 \Delta/h$ (panels (f--j)).
The frequency shift in the ground state $|g\rangle$ is first evaluated using Eq.~(\ref{fShift-Ground2}). All matrix elements are computed with a tight-binding model, as described in Appendix~\ref{app:TB}. We assume that scattering takes place only in the longitudinal direction, and hence does not mix the subbands. Thus, in absence of magnetic field the matrix elements of ${\cal H}'_0$ are then zero for all pseudospin-non-conserving transitions \cite{Park2017}. Frequency shifts in the other states are found from Eq.~(\ref{fShift-MB}). Transitions from $|g\rangle$ create pairs of excitations (pair transitions), leading for example (leftmost arrow in Fig.~\ref{Fig:Fig2}(c)) to the state  $\gamma^{\dagger}_{1d}\gamma^{}_{-1d}|g\rangle=-\gamma^{\dagger}_{1d}\gamma^{\dagger}_{1u}|g\rangle =-|1u1d\rangle$. Because of the redundancy between negative- and positive-energy states, we use here only labels corresponding to positive energies (Fig.~\ref{Fig:FigLongJunction}(a,f)). The states accessible from $|g\rangle$ and involving only the two lowest subgap levels are therefore those shown in Fig.~\ref{Fig:FigLongJunction}(b,g). We also consider the closest-in-energy states that can be reached from the single-particle state $|1u\rangle$ (Fig.~\ref{Fig:FigLongJunction}(c,h)). On the one hand, states with a single quasiparticle are accessible through single quasiparticle transitions, like $|2u\rangle=\gamma^{\dagger}_{2u}\gamma^{}_{1u}|1u\rangle$, or $|2d\rangle.$ On the other hand, the same fermion parity is also maintained with pair transitions that lead to states with three quasiparticles: $|1u1d2u\rangle=-\gamma^{\dagger}_{1d}\gamma^{}_{-2d}|1u\rangle,$ $|1u1d2d\rangle,$ and $|1u2u2d\rangle$. For all these possible states, frequency shifts are given by Eq.~(\ref{fShift-MB}), which simplifies to 
\begin{equation}
\delta f_r^{|\Psi\rangle} =  \delta f_r^{|g\rangle}+ \sum_{i>0,\sigma} n_{i\sigma} \delta f_r^{(i\sigma)} \;.\label{fShift-MBpos}
\end{equation}

Figure~\ref{Fig:FigLongJunction}(b,c,g,h) shows the transition energies from $|g\rangle$ and $|1u\rangle,$ with line colors encoding the resonator frequency shift for each transition (color scale in (d) or (e)). The phase asymmetry of the transition energies shown in panels (c) and (h) comes from the fact that we consider an initial state with a given pseudospin ($|u\rangle$). The mirrored spectra about $\delta=\pi$ would be obtained when considering transitions from $|1d\rangle$. 
The situation is the simplest when the resonator photons energy $hf_r$ is smaller than the energy of all the virtual transitions entering in the calculation of $\delta f_r^{(i\sigma)}$. All $\delta f_r^{(i\sigma)}$ are then dominated by the curvature term, and the resonator frequency shift for each transition is essentially related to the curvature of the transition energy. This is seen in Fig~\ref{Fig:FigLongJunction}(b,c) with the red color ($\delta f_r^{|\Psi\rangle}-\delta f_r^{|g\rangle/|1u\rangle}>0$) of the transition lines when they have positive curvature, blue for negative curvature. Detailed comparisons of the total shift with the curvature contribution are shown in Fig.~\ref{Fig:FigLongJunction}(d) for the pair transition $|g\rangle \rightarrow|1u1d\rangle$ and in Fig.~\ref{Fig:FigLongJunction}(e) for the single quasiparticle transitions $|1u\rangle \rightarrow|2u\rangle$ and  $|1u\rangle \rightarrow|2d\rangle.$ 

The results look more complicated in  Fig.~\ref{Fig:FigLongJunction}(g,h), with many sign inversions of the frequency shift when sweeping $\delta.$ Sign inversions occur when the energy of one of the virtual transitions entering in the calculation of the frequency shift in the initial or the final state coincides with $hf_r$. These coincidences are marked in panel (f) with small vertical bars linking levels with same spin distant by $hf_r$. For example, there is one of them at $\delta/\pi \approx 1.92,$ where $E_{2u}-E_{1u}=hf_r.$ Correspondingly, $\delta f_r^{(1u)}$ and $\delta f_r^{(2u)}$ present abrupt changes of sign at this phase, which is seen in all the lines involving $1u$ or $2u$ in Fig.~\ref{Fig:FigLongJunction}(g,h). Similarly, there is another such coincidence at $\delta/\pi \approx 1.27,$ where $E_{3d}-E_{2d}=hf_r,$ leading to color changes in the transition lines $|g\rangle \rightarrow |...2d\rangle.$ Detailed plots of the frequency shift for pair and single particle transitions are shown in Fig.~\ref{Fig:FigLongJunction}(i,j), with divergences when energy differences match $hf_r$.

\section{Gate and flux driving}
\label{gate-flux-drive}

As illustrated in Fig.~\ref{Fig:Fig1} transitions between Andreev states can be driven either with an ac electric field, using a gate \cite{Tosi2019}, or with a magnetic flux, by means of an ac current in a conductor placed  nearby the loop \cite{Hays2017}. The magnetic flux can also be modulated with an excitation applied through the resonator coupled to the superconducting loop \cite{Janvier2015}. The driving can be modeled by the following term added to the system Hamiltonian
\begin{equation}
\hat{A}(t) = \frac{1}{2} \sum_{i\sigma < j\sigma'} (A_{i\sigma, j\sigma'} \gamma^{\dagger}_{i \sigma} \gamma^{}_{j \sigma'} e^{i \omega_d t} + \text{h.c.}), 
\end{equation}
where $\omega_d = 2\pi f_d$ is the driving frequency. In the case of a flux driving, which acts on the phase $\delta$, 
$A_{i\sigma, j\sigma'} \propto 
\langle \Phi_{i\sigma}|{\cal H}'_0| \Phi_{j\sigma'}\rangle$.
In the absence of magnetic field and for a ballistic model which preserves the transverse spatial symmetry \cite{Park2017}, the ${\cal H}'_0$ operator does not mix the transverse channels of the weak link and thus only pseudospin-conserving transitions are allowed. Notice, however, that whenever the drive or the scattering breaks the 
transverse spatial symmetry spin-flip transitions can take place \cite{Hays2020}.

In the case of the gate driving the ac signal induces a displacement $\delta V(\vec{r})$ in the electrostatic potential experienced by the electrons in the junction region. The corresponding matrix elements in the driving Hamiltonian are thus
$A_{i\sigma, j\sigma'} = \langle \Phi_{i\sigma}|\delta V(\vec{r})\tau_z| \Phi_{j\sigma'}\rangle$, where $\tau_z$ is a Pauli matrix in electron-hole space.

In connection to recent experiments \cite{Tosi2019,Hays2017,Hays2019}, we analyse the case of spin-split Andreev states in semiconducting nanowires. 
As described in Ref.~\cite{Park2017}, the pseudospin of the Andreev states comes from nanowire's transverse modes with different spins hybridized by Rashba spin orbit coupling.
A perturbation $\delta V$ uniform in the transverse direction does not couple different transverse modes and therefore pseudospin flip transitions are not allowed, \textit{i.e.} $A_{iu, jd} = 0$. Only a non-uniform perturbation couples transverse modes and allows pseudospin flip transitions.

\begin{figure}[t]
\includegraphics[width=1\columnwidth]{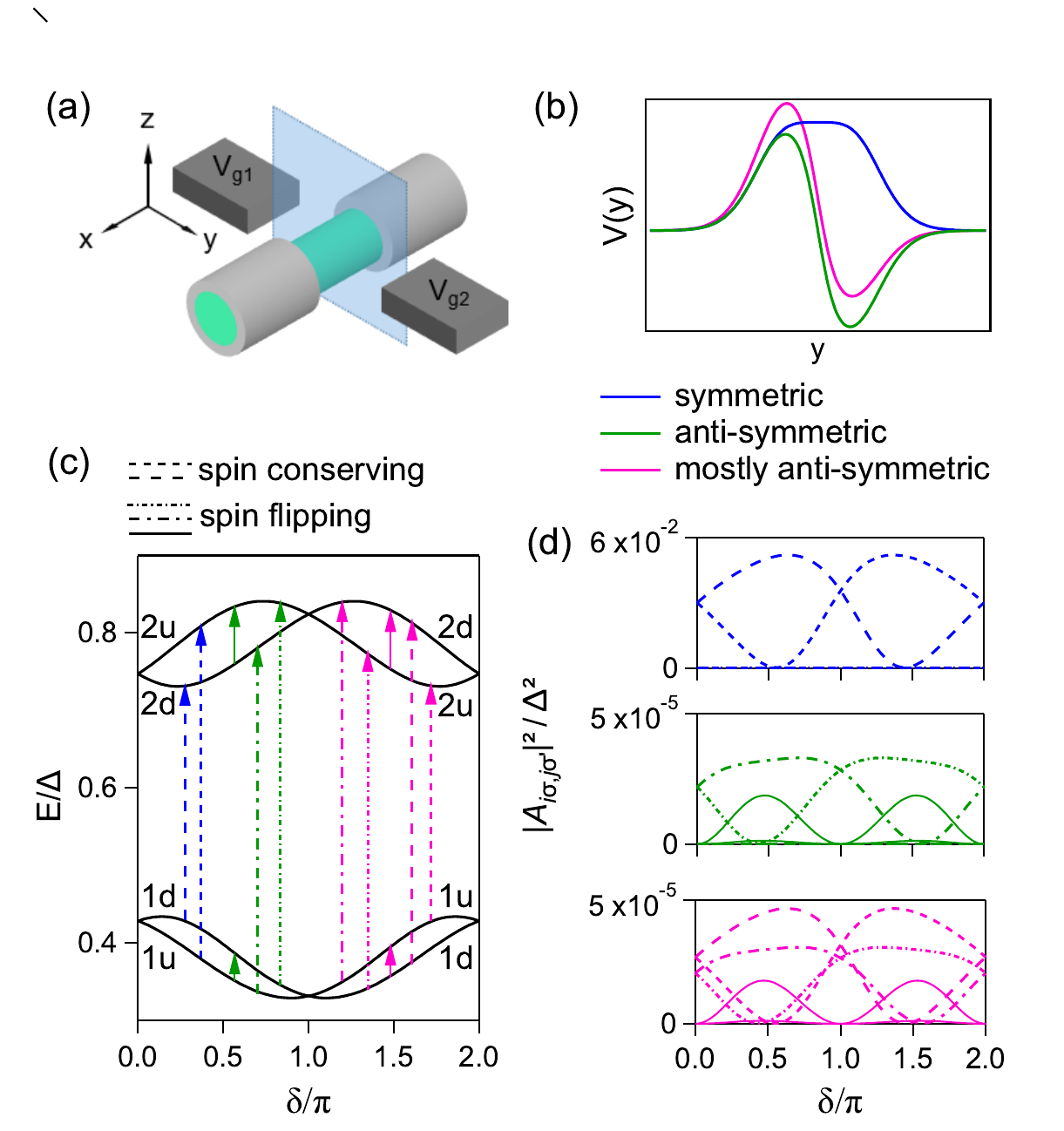}
\caption{(a) Schematic of a nanowire weak link with local side gates. (b) Induced driving potential in the transverse ($y$) direction of the nanowire in various situations:
symmetric (blue line), anti-symmetric (green line) or mostly anti-symmetric (magenta line) 
profiles can be obtained by controlling $V_{g1}$ and $V_{g2}$ applied to the gates. (c) Allowed transitions in the weak link with spin-split Andreev levels. Each color indicates the transitions induced by the driving potentials illustrated in (b). According to the corresponding matrix elements shown in (d) (calculated with the scattering model), the symmetric and anti-symmetric potentials lead to pseudospin-conserving (dashed arrows in (c), dashed lines in (d)) and pseudospin-flipping (dashed-dotted and full arrows in (c), dashed-dotted and full lines in (d)) transitions, respectively, and the mostly anti-symmetric potential results in both transitions with similar amplitudes. The square matrix elements for pseudospin-flipping intra-manifold transitions (solid lines) are $\sim 15$ times larger in the second manifold than in the first one, which at this scale is barely visible. 
}
\label{Fig:Fig7}
\end{figure}

The fact that all possible transitions between two Andreev manifolds have been observed in the experiments of Ref.~\cite{Tosi2019} indicates that the non-uniform component of the induced potential $\delta V$ by the gate electrode was significant.
More insight into the possibility to engineer the selection rules using gate driving can be obtained by considering the model of Ref.~\cite{Park2017} for the nanowire's transverse channels. Within this model the nanowire confining potential is assumed to have cylindrical symmetry. Thus, the modes in the lowest subband have zero angular momentum along the nanowire axis ($l=0$) and they have $l=1$ on the first excited subband. A lateral gate would impose a perturbation $\delta V(\vec{r})$ which typically breaks the rotational symmetry and therefore would couple states on different subbands, naturally leading to both pseudospin flip transitions and pseudospin conserving transitions. One could think, however, of a more general gate configuration like the one in Fig.~\ref{Fig:Fig7}(a), where two lateral gates can be set such that $\delta V(y) = -\delta V(-y)$ (or $\delta V(y) = \delta V(-y)$) in an anti-symmetric (or symmetric) configuration as indicated in the right panel. In the anti-symmetric case, the $\delta V$ matrix elements vanish for states on the same subband, but are finite for states in different subbands. As a consequence we would have $A_{iu,ju} = A_{id,jd}=0$. The allowed transitions between spin split Andreev states are indicated in Fig.~\ref{Fig:Fig7}(c) with arrows of different colors for symmetric (blue), anti-symmetric (green) or an intermediate (magenta) configuration. The phase-dependent matrix elements for each case are shown in Fig.~\ref{Fig:Fig7}(d) for the parameters that give the spectrum in (c) which correspond to a fit of the data in Fig.~\ref{Fig:FitPRX2} discussed below.

\section{Resonator frequency shifts for the driven weak link}
\label{driven-shift}

While a driven two-level system can be described using Bloch equations \cite{Palacios2010}, we develop here the theory for the general multi-level case in the weak drive regime. In order to obtain the frequency shift for the resonator coupled to the \textit{driven}
weak link we analyze the resonator spectral function $D^R(\omega) = -i\int_0^{\infty} dt e^{i \omega t} \langle [a(t),a^{\dagger}(0)]\rangle$. We take first the interaction picture in which the time evolution of the resonator and the weak link are provided by solving master equations including dissipation, and then we treat the resonator-weak link coupling and the drive as perturbations. We calculate the perturbation terms up to second order in both $\delta_{\text{zp}}$ and $A_{i\sigma, j\sigma'}$ (see Appendix~\ref{app:Spectr} for details).

The frequency shift for a single quasiparticle transition from $|i_0 \sigma_0\rangle$ is
\begin{equation}
\delta f_r^{\text{SQPT}} = 2 \sum_{j\sigma > 0	} \frac{|A_{i_0\sigma_0,j\sigma}|^2}{|D_{i_0\sigma_0,j\sigma}|^2} \left( \delta f_r^{(j \sigma)} - \delta f_r^{(i_0 \sigma_0)}\right),\label{SQPT}
\end{equation}
where $D_{a,b} = \hbar \omega_{d} - |E_{a} - E_{b}| + i (\Gamma_{a}+\Gamma_{b}) \hbar/2$, 
$\omega_{d}$ is the driving field frequency and $\Gamma_{a (b)}$ are phenomenological parameters to account for the finite linewidths in the transition spectrum, which in our present theory are associated to the states relaxation.

For a pair transition from the ground state $|g\rangle$, one obtains 
\begin{equation}
\delta f_r^{\text{PT}} = 2 \sum_{\{j\sigma, k\sigma'\}}  \frac{|A_{-j\bar{\sigma},k\sigma'}|^2}{|D_{-j\bar{\sigma},k\sigma'}|^2} \left( \delta f_r^{(j \sigma)} + \delta f_r^{(k \sigma')}\right), \label{PT}
\end{equation}
where $\{j\sigma, k\sigma'\}$ means a set of indices $j\sigma$ and $ k\sigma'$ corresponding to positive energy levels ordered in energy, and does not contain a permutation of the indices. $j\sigma$ and $-j\bar{\sigma}$ are for a pair of particle-hole symmetric Andreev levels. 

\begin{figure*}[t]
\includegraphics[width=2\columnwidth]{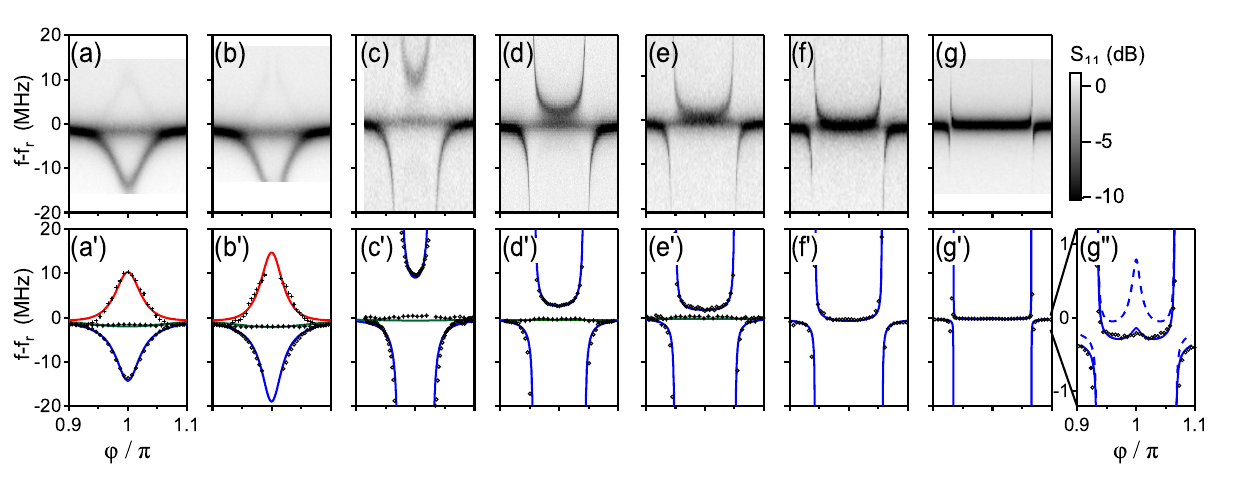}
\caption{Fit of single-tone continuous-wave spectroscopy data taken with a vector network analyzer (VNA) on a series of atomic contacts obtained from the same sample (described in  Ref.~\cite{Janvier2015}), with resonator at $f_r=10.1~$GHz . (a-g) Raw data, with reflection coefficient of the resonator $S_{11}$ coded with gray scale. By fitting $S_{11}(f)$ at each flux with the sum of shifted resonance curves, up to three values of the frequency shift could be extracted ((a'-g') and (g"), symbols). They are associated with the shifts in states $|g\rangle$, $|o\rangle$ and $|e\rangle$ of the channel with the largest transmission. Solid line are fits with complete theory, using $\delta_{\text{zp}}=0.0042$ and $\Delta/h=44.3~$GHz. Blue: $\delta f_r^{|g\rangle}$, green: $\delta f_r^{|o\rangle}$ (with respect to 1st channel), red in (a',b'): $\delta f_r^{|e\rangle}$. Fit parameters are given in Appendix~\ref{app:Transmissions}. In (g''), the dashed line corresponds to $\delta f_r^{|g\rangle}$ in the rotating wave approximation (Eq.~\ref{RWA}). }
\label{Fig:fitAC}
\end{figure*}

\section{Comparison with experiments}
\label{comparison-AC}
\subsection{Atomic contacts}

We first focus on the simplest experiments, reported partly in Ref.~\cite{Janvier2015}, dealing with atomic contacts hosting a small number of transport channels. The superconducting loop containing the atomic contact is coupled to a microwave resonator at $f_r=10.1$~GHz measured in reflection. In the experiment, atomic contacts with various channel transmissions are formed and probed with the same sample.

We first discuss single-tone continuous-wave (CW) spectroscopy data taken on seven different atomic contacts, as shown in Fig.~\ref{Fig:fitAC}. To acquire these data the microwave response of the resonator is probed as a function of flux $\varphi=2\pi \Phi/\Phi_0$, with no drive applied on the weak link. Over a small flux range around $\varphi=\pi$, the amplitude of the reflection coefficient $|S_{11}|$ 
displays up to three distinct local minima (in dark) as a function of frequency, as shown in Fig.~\ref{Fig:fitAC}(a-g). The positions $f_{1,3}$ of these minima were extracted by fitting $|S_{11}|(f)$ with the linear combination $\sum_{i=1}^3 p_i |S_{11}^0|(f,f_i)$ of resonance lines $|S_{11}^0|(f,f_i)$ corresponding to a single resonance centered at $f_i$. The extracted $f_i$ are shown with symbols in Fig.~\ref{Fig:fitAC}(a'-g').
These data can essentially be understood by considering the contribution of just one dominant channel of transmission $\tau_1$ such that the corresponding Andreev frequency $f_{A1}$ comes very close to the resonator frequency $f_r$ (a,b) or crosses it (c-g). The data are ordered with increasing $\tau_1$ from (a) to (g). The three resonances are attributed to partial occupancy of ground, odd and excited states for the corresponding channel. As shown in Eqs.~(\ref{AC-fShift-Ground2},\ref{AC-fShift-excited}), and since, according to the analysis of Section~\ref{nodrive}, contributions of the continuum can safely be neglected for phases close to $\pi$, the frequency shifts associated with ground and excited state are opposite (Fig.~\ref{Fig:fitAC}(a,b)), and the frequency shift associated with the odd state is zero (Fig.~\ref{Fig:fitAC}(a,e)). Finally, in order to explain a  small residual  global shift, it is necessary to consider the contribution of one or two additional channels with smaller transmissions, as explained below.

\begin{figure}[t]
\includegraphics[width=0.8\columnwidth]{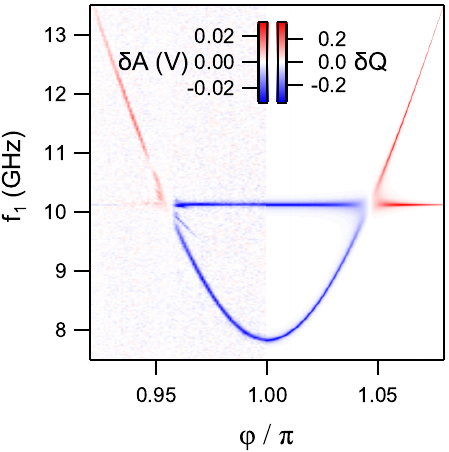}
\caption{Two-tone spectrum measured on the same atomic contact as for Fig.~\ref{Fig:fitAC}(d). Left: experimental data. Color scale corresponds to the change of amplitude of a quadrature of the reflected signal. Right: calculated spectrum, line color codes change in $S_{11}$, scaled in order to fit at best the data.}
\label{Fig:AC2tone}
\end{figure}

The resonator frequency shift is obtained from Eqs.~(\ref{AC-fShift-Ground2},\ref{AC-fShift-odd},\ref{AC-fShift-excited}), adding the contributions of all channels. The bare resonator frequency $f_r$ was determined from measurements of the resonator with open contacts. The phase $\delta$ across the contact differs slightly from $\varphi=2\pi \Phi/\Phi_0$ due to the phase drop across the loop inductance (see Appendix~\ref{app:screening}).
In a first step, the value of $\delta_{\text zp}=0.0042$, common to all contacts, was determined by fitting Fig.~\ref{Fig:fitAC}(d), taken on a contact for which a fit of the two-tone spectroscopy yielded $\tau_1=0.992$  (\cite{Janvier2015} and Fig.~\ref{Fig:AC2tone}). For the other contacts, the fitting parameters are the transmissions $\tau_{i}$ of two or three channels (fit parameters given in Appendix~\ref{app:Transmissions}). The transmission $\tau_1$ of the most-transmitted channel determines the overall shape of the spectra. In the small phase interval considered here, the effect of the other channels is simply  a constant overall shift of the order of a MHz. The predictions for $\delta f_r$ assuming the most-transmitted channel to be in $|g\rangle$, $|o\rangle$ or $|e\rangle$ are shown with blue, green and red lines.

The $\delta-$ and $\tau-$ dependence of the coupling constant $g$ is an essential ingredient to obtain a consistent fit of all the data at once. The difference between the full theory and the JC contribution is very small at the scale of Fig.~\ref{Fig:fitAC}. In Fig.~\ref{Fig:fitAC}(a--c), since $|f_{A1}(\pi)-f_r| \ll f_{A1}(\pi)+f_r,$ the counter-rotating term $\propto 1/(f_{A1}+f_r)$ can also be neglected, and the RWA is sufficient. When $\tau_1$ is closer to 1 (Fig.~\ref{Fig:fitAC}(d-g)), $f_{A1}$ becomes significantly smaller than $f_r$ near $\delta=\pi$ and the counter-rotating term must be taken into account, as illustrated in Fig.~\ref{Fig:fitAC}(g'') where the RWA prediction shown with a dashed line departs clearly from the data.

We present in Fig.~\ref{Fig:AC2tone} the two-tone spectroscopy measured for the atomic contact of Fig.~\ref{Fig:fitAC}(d)   \cite{Janvier2015}. A single transition appears in the measurement window, corresponding to a channel with transmission $\tau \approx 0.992.$ To acquire this data a strong microwave pulse drives the two-level system, during a time exceeding its relaxation and dephasing times, immediately followed by a microwave tone probing the resonator. The two  quadratures of  the signal reflected by the resonator are measured by homodyne detection. In Fig.~\ref{Fig:AC2tone}, the color scale represents the change of amplitude $\delta A$ of one of these quadratures, relatively to its value in the absence of excitation. It depends on the steady state occupancy of the states after the excitation pulse and on the frequency shift of the resonator in each state. For this data taken on a very high-transmission contact and around $\delta=\pi$, the resonator frequency shifts are dominated by the dispersive shifts, with negligible effects of the states in the continuum, and the rotation wave approximation (Eq.~(\ref{RWA})) applies: $\delta f_r^{|g\rangle}\simeq-\delta f_r^{|e\rangle}\simeq -\left(g(\delta)/2\pi\right)^2/(f_A-f_r)$ with $g(\delta)\simeq g(\pi)  E_A(\pi)/E_A(\delta)$ (from Eq.~(\ref{AC-matelem}), when $E_A \ll \Delta$). 
When the resonator frequency is $f_r+\delta f_r$, the complex reflection coefficient of a measurement tone at frequency $f_r+\delta f_m$ on the resonator is \cite{Janvier2015} 
\begin{equation}
    S_{11}(\delta f_r)=1-\left(1+\mathrm{e}^{i \theta}\right)/(1+Q_{\text{ext}}/Q_{\text{int}}) 
    \label{Eq.:S11}
\end{equation}
with
\begin{equation}
    \theta=-2 \arctan \left(2 Q_t (\delta f_r-\delta f_m)/f_r \right).
\end{equation} 
In this expression, $Q_{\text{int}},$ $Q_{\text{ext}}$ and $Q_t$ are the internal, external and total quality factors of the resonator. 
When driving the system, if it is in an even state (probability $1-p_o$), the occupancies of the ground and excited states change by $\delta p_g$ and $\delta p_e=-\delta p_g$, resulting in a change of $S_{11}$ by $\delta S_{11}=\delta p_g (S_{11}(\delta f_r^{|g\rangle})-S_{11}(\delta f_r^{|e\rangle})).$ Putting everything together, one gets, for $\delta f_m=0,$
\begin{equation}
    \delta S_{11}=\delta p_g \frac{8i}{1+\frac{Q_{\text{ext}}}{Q_{\text{int}}}} \frac{u}{1+4 u^2} =i \delta Q,
    \label{eq:Bloch}
\end{equation}
with $u=Q_t \delta f_r^{|g\rangle}/f_r$.
To compute $\delta p_g,$ one uses the result of the Bloch equations  \cite{Palacios2010} adapted for the presence of the odd state:
\begin{equation}
    \delta p_g=\frac{p_e^0-\frac{1-p_o}{2}}{1+\frac{1+(T_2 \delta \omega)^2}{T_1 T_2 \omega_R^2}},
\label{deltapg}
\end{equation}
with $T_1$ and $T_2$ the life time and coherence time, $\delta \omega=2\pi (f_1-f_A)$ the detuning between the drive frequency $f_1$ and the qubit frequency $f_A$, and $\omega_R$ the Rabi frequency. For simplicity, we assume here that $T_1$ and $T_2$ are constant: $T_1=4~\mu$s and $T_2=38~$ns (values measured at $\delta=\pi$). Since the excitation acts on the phase across the contact, the Rabi frequency depends on $\delta$: $\omega_R \propto {\cal M}^2$ (see Eq.~\ref{AC-matelem}). The drive tone being sent through the resonator, its amplitude is filtered: $\omega_R \propto (1+Q_\text{t}^2(f_1/f_r-f_r/f_1)^2)^{-1/2}.$  In the fitting of the data, $\omega_R$ is set to $2\pi \times 4.2~$MHz at $\delta=\pi$ in order to reproduce the measured line width.

Using  $Q_t=2200,$ $Q_\text{int}=4800,$ $f_r=10.13~$GHz, and $g(\pi)/2\pi=72~$MHz, one obtains the fit shown on the right hand side of Fig.~\ref{Fig:AC2tone}, with the color scale of $\delta Q$ adapted to match the data. Not only does the change in the $Q$ quadrature reproduce the changes $\delta A$ in the measured quadrature on the resonance line $f_A(\delta),$ but one also predicts a signal at $f_1=f_r,$ which has its origin in the very large Rabi frequency when the drive signal is not filtered by the resonator. At this precise frequency, the strong detuning $\delta \omega$ is  compensated by the large $\omega_R$ in Eq.~(\ref{deltapg}), and  $\delta p_g$ is non-zero even if the drive is far from the resonant frequency $f_A$. This feature is clearly visible in the data, although not as strong as in the calculation for $f_A>f_r$ perhaps due to an effect of the resonator non-linearity not included in the model.

\begin{figure}[th!]
\includegraphics[width=\columnwidth]{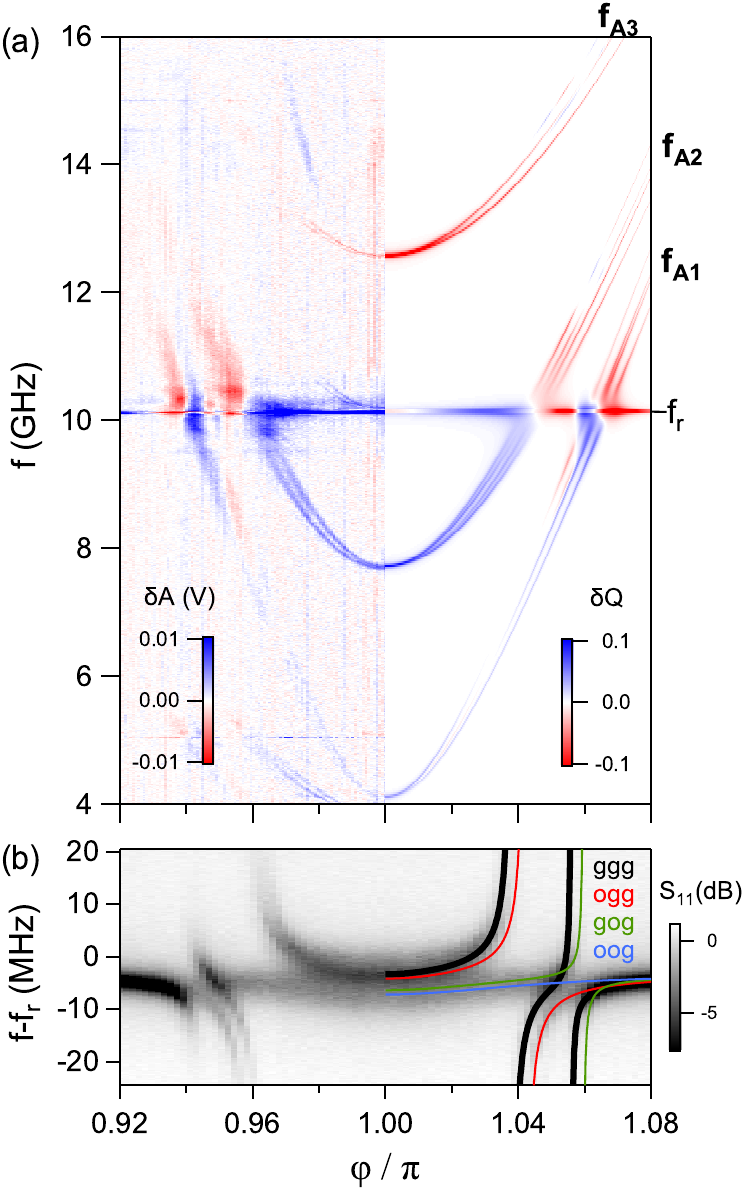}
\caption{(a) Two-tone spectroscopy of an atomic contact. On the right half of the figure, calculated spectrum (see text) with identification of three Andreev transitions with frequencies $f_{A1,2,3}$. Lines color corresponds to the calculated change in $Q$ quadrature of $S_{11}$.
(b) Grayscale codes reflection coefficient amplitude $|S_{11}|$. Lines on the right-hand side are theoretical resonator shifts depending on the states occupancy: black (ggg): all channels in ground state; Red: first channel in odd state (ogg);
Green: 2nd channel in odd state (gog); Blue: first and second channel in odd state (oog). 
A global shift of -3.4~MHz was applied to the theory curves, which can be attributed to the effect of several low-transmitting channels that are not visible in the two-tone spectrum.}
\label{Fig:fitAC2}
\end{figure}

\begin{figure}[h!]
    \centering
    \includegraphics[width=0.8\columnwidth]{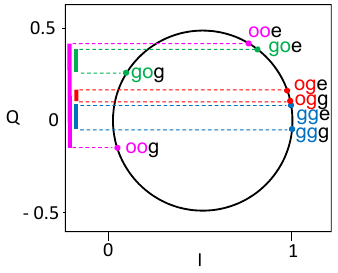}
    \caption{ Representation in the $IQ$ plane of $S_{11}$ at $\varphi=1.04\pi,$ for the states involved in transitions on the third channel (see dashed line in Fig.~\ref{Fig:fitAC2}(a)). Depending on the state of the two first channels, the change of $Q$ takes very different values, the largest change corresponding to $|oog\rangle \rightarrow |ooe\rangle$ (in magenta). In Fig.~\ref{Fig:fitAC2}(a), $\delta Q$ for the four transitions is further scaled by the probability of each initial state, \textit{e.g.} $p_{o1}p_{o2}(1-p_{o3})$ for $|oog\rangle \rightarrow |ooe\rangle$. Among the four transition lines, only this one is clearly visible in Fig.~\ref{Fig:fitAC2}(a).}
    \label{fig:S11polar}
\end{figure}

In Fig.~\ref{Fig:fitAC2}, we show the two-tone spectroscopy and the single-tone CW data of one particular contact obtained with the same sample and showing a double avoided crossing. In the two-tone spectroscopy (Fig.~\ref{Fig:fitAC2}(a)), one observes three Andreev transitions (labeled $f_{A1}$, $f_{A2}$ and $f_{A3}$) corresponding to channels with transmissions $\tau_1=$0.998, $\tau_2=$0.992 and $\tau_3=$0.980, with minimum transition frequencies (at $\delta=\pi$) of 4.1, 7.7, and 12.6~GHz. Two transition lines $f_{A1}$ and  $f_{A2}$ cross the resonator at 10.1~GHz. 
The experimental data (left half of Fig.~\ref{Fig:fitAC2}(a)) show split transition lines at $\delta \neq \pi.$ This is accounted for by the relation between the phase $\delta$ across the contact and the reduced flux $\varphi,$ which involves the phase drop across the loop inductance, proportional to the current through the atomic contact (see Appendix~\ref{app:screening}). In the presence of several channels which can be either in the ground or the odd state (we neglect the occupancy of the excited states), this current can take several values. As a consequence, the phase across the contact, common to all the channels, depends on the states occupancy, which gives rise to an effective coupling between the channels.
To calculate the spectrum from theory, we use Eq.~(\ref{Eq.:S11}) (with $\delta f_m=-0.4~$MHz) for each state $|\Psi\rangle$, and compute the weighted average $S_{11}=\sum_{|\Psi\rangle} p_{|\Psi\rangle}S_{11}(|\Psi\rangle)$ using the probabilities $p_{|\Psi\rangle}$ for each state. The probabilities to find each channel in the odd state were taken constant, $p_{o1}=0.55,$ $p_{o2}=0.5,$ $p_{o3}=0.4,$ determined by fitting the phase dependence of $S_{11}(\phi)$ at an excitation frequency where no transition is observed ($f_1=16~$GHz). In presence of the excitation pulse, the change in each $p_{|\Psi\rangle}$ was obtained from Eq.~(\ref{eq:Bloch}). For simplicity, we assumed $T_1=4~\mu$s and $T_2=38~$ns at each phase and for each state. Depending on the state of the other channels, a transition from $|g\rangle$ to $|e\rangle$ in one channel leads to different changes in $S_{11}$, as illustrated in Fig.~\ref{fig:S11polar}. The resulting calculated spectrum is shown in the right half of Fig.~\ref{Fig:fitAC2}(a), in which we represent the changes $\delta Q$ in the $Q$ quadrature.  Fainter lines in the data are multi-photon transitions (at $f_{Ai}/n$), or transitions involving two channels ($f_{A1}+f_{A2}$ and $(f_{A2}+f_{A3})/2),$ not included in the theoretical plot.
Despite the strong simplifications in the analysis, the changes $\delta A$ in the amplitude of the reflected signal in Fig.~\ref{Fig:fitAC2}(a) are well reproduced by the calculated changes in the $Q$ quadrature of $S_{11}$. Similar agreement is found on the other quadrature.

In Fig.~\ref{Fig:fitAC2}(b) we show the single-tone spectroscopy of the resonator. The resonator frequency shift calculated under the assumption that the three channels are in the ground state is shown as a black solid line. Features associated with configurations in which one (red and green lines) or both (blue line) of the two most-transmitted channels are in the odd state are also observed. 
Horizontal shifts of the red and green lines with respect to the black ones result from the phase drop across the loop inductance, different in each configuration.
The analysis of the data of Fig.~\ref{Fig:fitAC2} illustrates how, in a multi-channel weak link, the frequency shift associated with a transition in one channel  depends strongly on the occupancy of the Andreev states in the others.

\subsection{Nanowire junctions}
\begin{figure}[t]
\includegraphics[width=0.8\columnwidth]{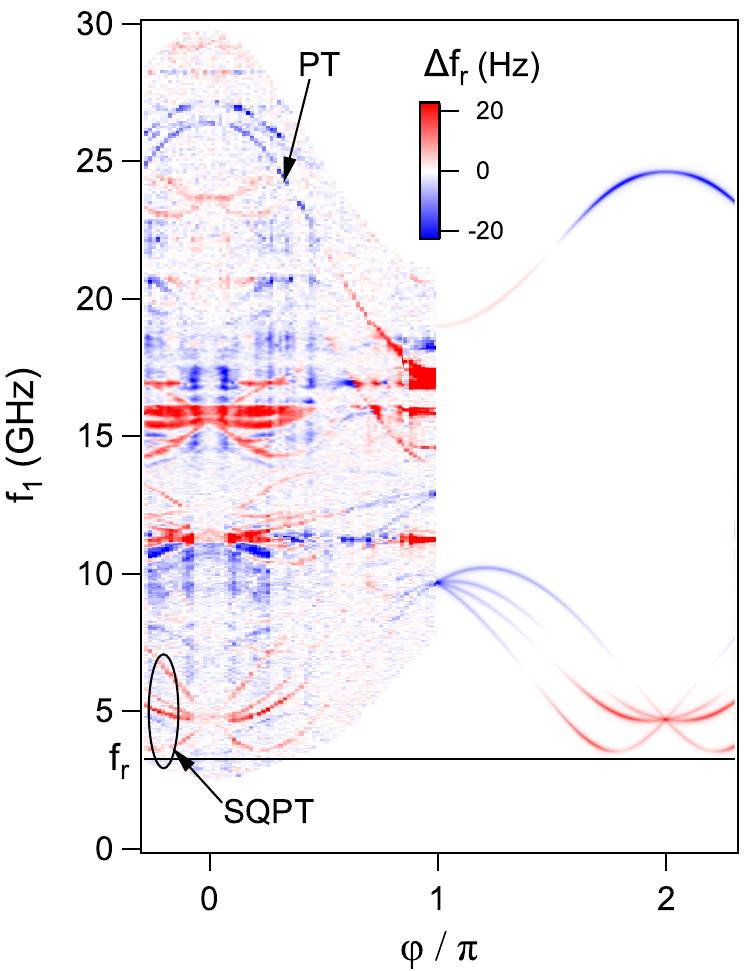}
\caption{Two-tone spectroscopy of an InAs nanowire weak link (from Ref.~\cite{Tosi2019}). Left: experimental data. Colorscale represents the resonator frequency shift (sign corrected compared to Ref.~\cite{Tosi2019}). A pair transition (PT) and a bundle of single particle transitions (SQPT) are pointed at. Right: calculation for a single occupied channel (see text). Colorscale is the difference in frequency shift between initial and final state. Solid line at 3.26~GHz indicates the resonator frequency. In the calculation, it was assumed that $\delta_{\text{zp}}= 1.2 \times 10^{-5}$ and the dissipation rate is $\Gamma_{1\sigma}+\Gamma_{2\sigma'}=0.62\, \text{GHz}$. 
The sign of the frequency shift in the experiment has been corrected compared to Ref.~\cite{Tosi2019}.}  
\label{Fig:FitPRX1}
\end{figure}

\begin{figure}[t]
\includegraphics[width=0.8\columnwidth]{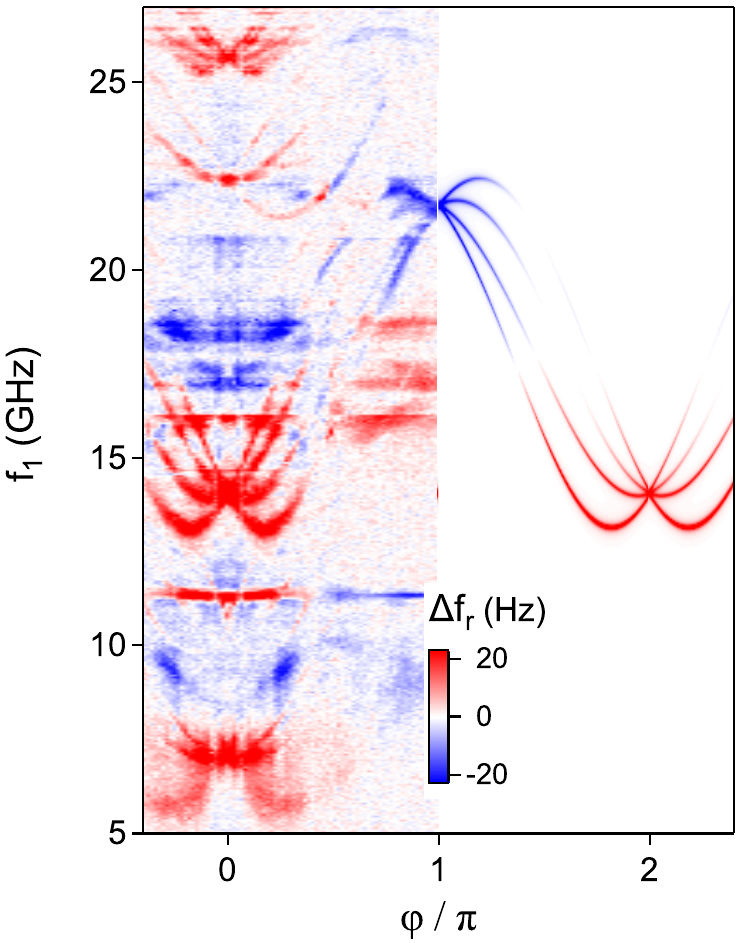}
\caption{Comparison between experimental results on Andreev transitions in a nanowire junction \cite{Tosi2019} and a full calculation of the shift taking into account that the transitions were induced by microwaves applied on the gate. The parameters for the calculation are those that allowed fitting the spectrum \cite{Tosi2019} and the matrix elements for the microwaves shown in Fig. \ref{Fig:Fig7}(c) with magenta are used. The same values of $\delta_{\text{zp}}$ and the dissipation rate are used as in Fig.~\ref{Fig:FitPRX1}.}  
\label{Fig:FitPRX2}
\end{figure}

The two-tone spectra shown in the left-half of Figs.~\ref{Fig:FitPRX1} and \ref{Fig:FitPRX2} were measured on the same InAs nanowire device at two different gate-voltages \cite{Tosi2019}. In this experiment, the coupling between the weak link and the resonator was two orders of magnitude smaller than in the atomic contacts case, and consequently  the resonator frequency shift was always smaller than its linewidth. In this limit, $\Delta S_{11}$ depends linearly on the sum of the frequency shifts associated to each level, and the complications of the well-coupled system described in the previous section can be ignored. Hence, the color scales in Figs.~\ref{Fig:FitPRX1} and \ref{Fig:FitPRX2} directly corresponds to the resonator frequency shift. Both spectra correspond, like in  Fig.~\ref{Fig:FigLongJunction}(a-e), to a situation in which the resonator frequency $f_r=3.26$~GHz is very low as compared to most of the observed transition lines.

A pair transition and several single-particle transitions are clearly recognized in the spectrum of Fig.~\ref{Fig:FitPRX1}. The signal at each point being integrated over 100~ms, a time much longer than the one for parity changes due to quasiparticle poisoning \cite{Hays2017}, transitions from states of different parities appear in the spectrum. From the analysis illustrated by Fig.~\ref{Fig:FigLongJunction}(a-e), one understands that the frequency shifts corresponding to transitions above the resonator frequency are essentially given by the curvature of the transition lines: with the color scale of Fig.~\ref{Fig:FitPRX1}, lines are red where they have positive curvature, and blue when negative.
For a more quantitative comparison with theory, we first fit the position of the bundle of single-particle transitions appearing in the range 3-10~GHz, using the scattering model presented in Ref.~\cite{Tosi2019} (see parameters in Appendix~\ref{app:spiderfits}). 
The calculated single particle transition lines shown in the right-half of Fig.~\ref{Fig:FitPRX1} reproduce well the observed single particle transition energies, but the pair transition predicted from the same Andreev levels disperse less than in the experimental data.
Using these parameters, we calculate the matrix elements for ${\cal H}_0'$ (see Appendix B). Since the two-tone spectroscopy data was taken at very small power, the theory of weak-driving is fully applicable. We evaluate the matrix elements for the weak-driving through the gate and compute the resonator shift shown in the figure, using Eqs.~(\ref{SQPT},\ref{PT}).
Globally, the shifts calculated for the four single particle transitions reproduce quite well the observed ones. However, some details differ, notably for the highest single quasiparticle transition, with shifts near $\varphi=0$ larger in the data than in the calculation. The shift for the pair transition is reproduced only at a qualitative level.

A similar procedure was used to fit the data in Fig.~\ref{Fig:FitPRX2}, taken at another value of the gate voltage \cite{Tosi2019}, but again in a situation where the effect of states' curvature dominates. The fit parameters are given in Appendix \ref{app:spiderfits}. In this case, the calculated pair transition lies outside the frequency range of the graph and only the bundle of four single-particle transitions is clearly recognisable (transitions in the range 13-23~GHz). Theory captures most of the features of the experiment for this set of single particle transition lines.

\begin{figure}[t]
\includegraphics[width=\columnwidth]{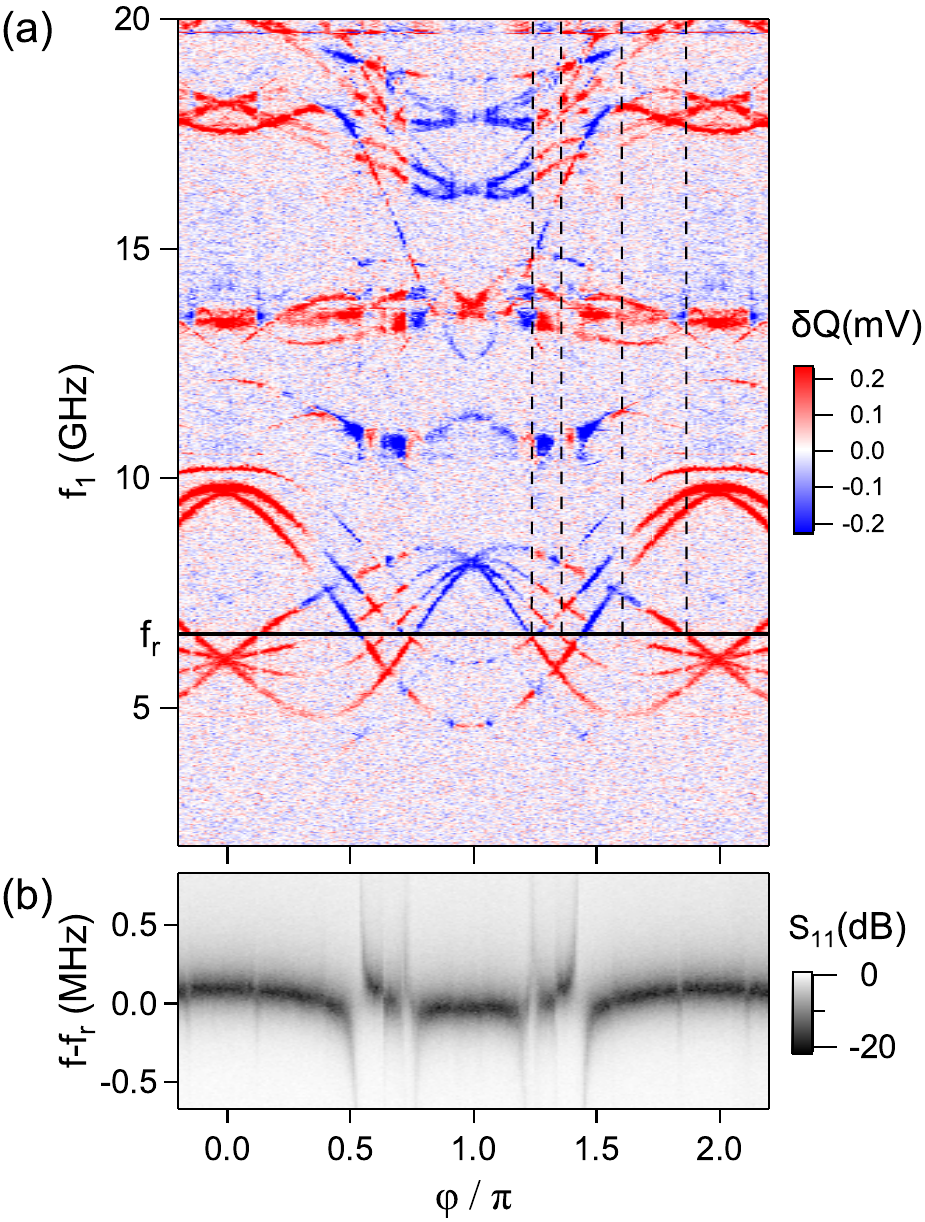}
\caption{(a) Two-tone spectrum measured on an InAs nanowire weak link of length $L\sim 550$~nm, and using a resonator at $f_r=6.6$~GHz. The color-coded quadrature of the measured signal shows many sign changes along the transition lines, qualitatively in agreement with the behavior illustrated in Fig.~\ref{Fig:FigLongJunction}(g,h): the sign changes are attributed to situations where the energy of some virtual transitions match $h f_r.$ (b) Associated single-tone spectrum.}
\label{Fig:newspec}
\end{figure}

\begin{figure}[t]
\includegraphics[width=\columnwidth]{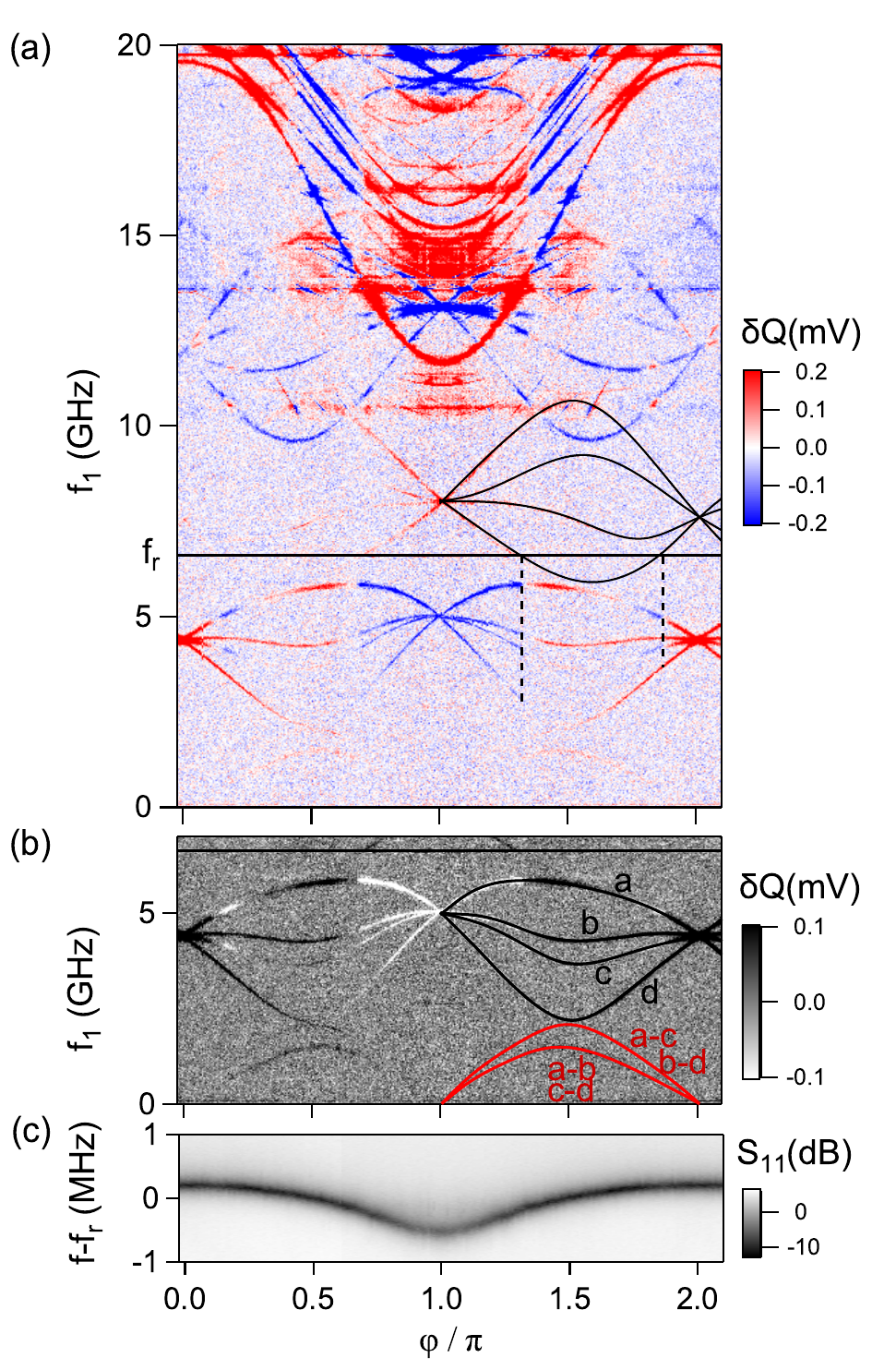}
\caption{(a,b) Two-tone spectrum measured on an InAs nanowire weak link of length $L\sim 550$~nm (different gate tuning compared to previous one), and using a resonator at $f_r=6.6$~GHz. The color-coded quadrature of the measured signal shows sign changes along the transition lines, qualitatively in agreement with the behavior illustrated in Fig.~\ref{Fig:FigLongJunction}(g,h): the sign changes are attributed to situations where the energy of some virtual transitions match $h f_r.$ For example when the lowest transition line of the second group of single particle transitions (underlined with black splines in (a)) crosses the resonator, the sign of frequency shift along the transition lines in the lowest group of SQPT changes. In (b), same data as (a) but stronger contrast and other colorscale, intra-manifold spin-flip transitions are visible. The red lines that superimpose on the data are obtained as differences between the inter-manifold transition energies underlined in black and labeled a,b,c,d. (c) Single-tone spectrum.}
\label{Fig:newspec2}
\end{figure}

In the experimental results reported in Ref.~\cite{Tosi2019}, the resonators shift was remarkably low (tens of Hz) as compared to that observed for atomic contacts (tens of MHz) \cite{Janvier2015}.
There are two reasons for this. On the one hand, the geometry of the circuit, which determines the phase fluctuations the resonator induces in the loop. It can be optimized with the circuit design. On the other hand, and more fundamental, the reduction of the matrix element of ${\cal H}_0'$ in the long-junction limit. As a rough approximation (more accurate close to $\delta=\pi$), the matrix element expression reduces to that for a short junction with an effective gap $\Delta_{\text{eff}}=\frac{\Delta}{1+\lambda}$; this contributes to a $(1+\lambda)^2$ reduction in the coupling. 

\begin{figure}[t!]
\includegraphics[width=1\columnwidth]{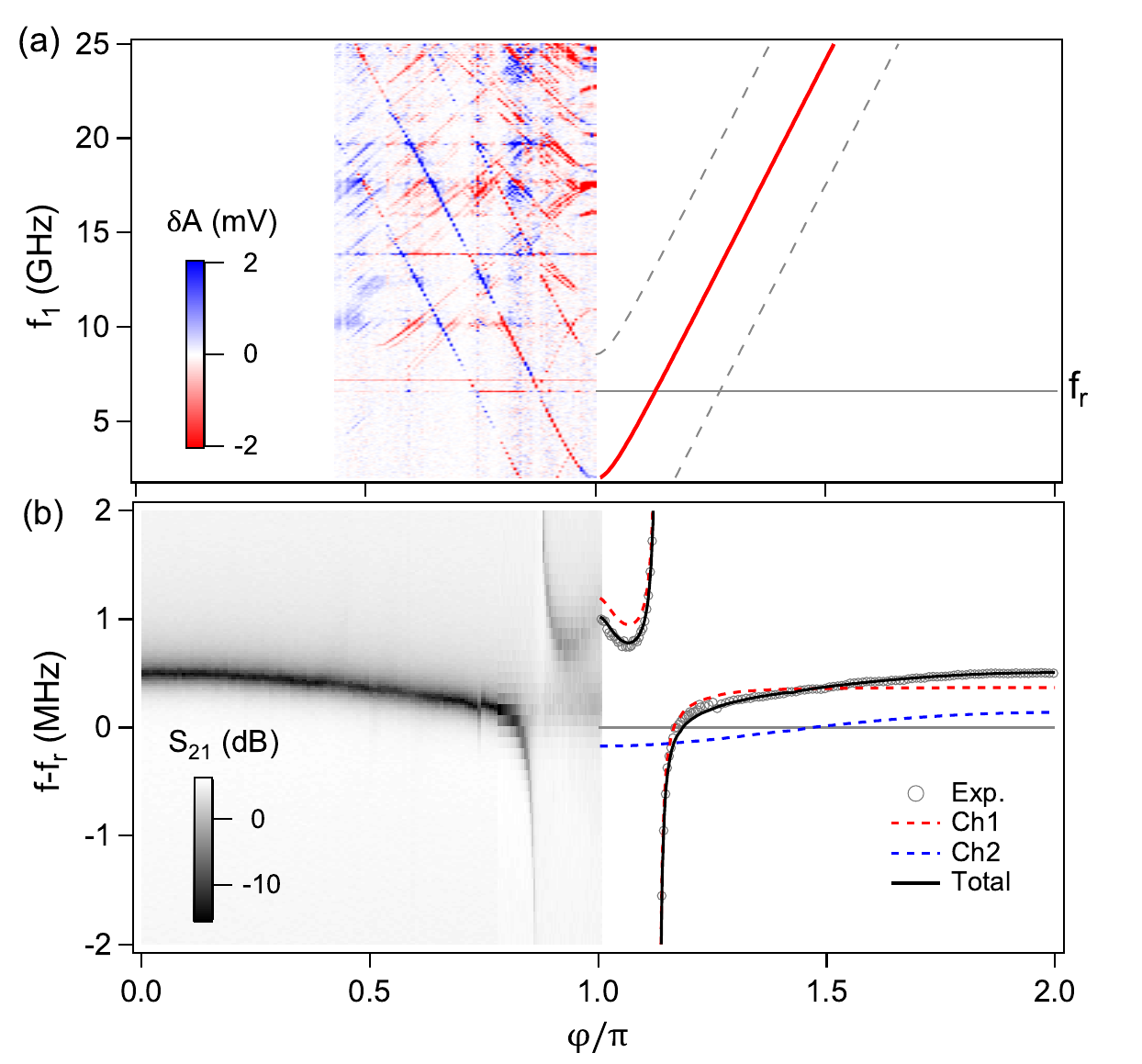}
\caption{(a) Two-tone spectroscopy of an InAs nanowire weak link of length $L\sim 550$~nm, showing a pair transition $f_A$ anticrossing the resonator at $f_r=6.60762$~GHz. Red line on the right-side is a fit of this transition (see Appendix~\ref{app:spiderfits}) with its two replicas at $f_A\pm f_r$ shown in dashed gray. (b) Single-tone spectroscopy. (Left) Transmission coefficient amplitude $S_{21}$ measured with vectorial network analyzer. (Right) Comparison with theory. For each phase value, the resonator frequency is extracted from the raw data ; the shift from its bare value is shown with gray disks and compared to the calculated shift due to a single channel (dashed red), whose pair transition towards the lowest Andreev manifold fits the transition line $f_A$ shown in (a). Blue dashed line: contribution of an effective second channel (parameters in Appendix~\ref{app:spiderfits}; because of its low transmission, the corresponding pair transition does not fall in the frequency range of (a)). Black line: total shift due to both channels.}
\label{Fig:Fig11}
\end{figure}

In recent experiments \cite{Metzger2021} we coupled the nanowire through a shared inductance, increased the resonator frequency and diminished its impedance, leading to an enhanced  $\delta_{\text{ZP}}$ by three orders of magnitude. The data from Figs.~\ref{Fig:newspec}, \ref{Fig:newspec2} and \ref{Fig:Fig11} have been taken using this new setup. The sample comprises a $L\sim 550$~nm-long weak link from a full shell InAs-Al nanowire  (same batch as the one in Ref.~\cite{Tosi2019}), coupled to a resonator with $f_r=6.6$~GHz. The coupling being much larger than in Figs.~\ref{Fig:FitPRX1} and \ref{Fig:FitPRX2}, the measured signal cannot be easily converted in a resonator shift, and we show the change in one quadrature, $\delta Q$.

In the range between 5 and 10~GHz of the otherwise very busy spectrum of Fig.~\ref{Fig:newspec}, one recognizes three pair transitions and a bundle of single particle transitions. The color-coded quadrature shows many abrupt sign changes along the transition lines, like the behavior illustrated in Fig.~\ref{Fig:FigLongJunction}(g,h): the sign changes are attributed to situations where the energy of some virtual transition matches the resonator frequency (as indicated by the black dashed lines). 

This spectrum has other remarkable features, like the occurrence of sets of pair transitions very close in energy. This is something we have observed many times in these nanowire weak links (see for example Fig. \ref{Fig:newspec2} between 11-20~GHz). Although we do not have a closed explanation yet, it could originate from subband orbital quasi-degeneracies due to the approximate rotational symmetry of the weak link, or to multiple subbands occupancy.
Moreover, the shape of several other lines in the spectrum does not correspond to what our simplistic model predicts, and levels anticrossings might be required to explain them.

In Fig.~\ref{Fig:newspec2} we show data measured in a different cooldown. 
In this case, changes of sign of the displayed quadrature occur when the lowest transition of a single-particle transition bundle (drawn in black based on the signal of both quadratures) crosses the resonators line. The resonator shift measured in the CW single-tone spectroscopy (Fig.~\ref{Fig:newspec2}(c)) shows a dominant contribution from the pair transition that lies at 12~GHz at $\delta=\pi$, but the anti-crossings expected at the position of the dotted lines in Fig.~\ref{Fig:newspec2}(a)
are not visible, indicating the very small occupancy of the initial state for the corresponding transition. This is in agreement with the difference in intensity between the transition lines in the two-tone spectroscopy. 
A remarkable feature in this spectrum is the presence of two very low-frequency lines (below 2~GHz), better seen in Fig.~\ref{Fig:newspec2}(b).
By analyzing their position in energy, they can be identified as spin-flip intra-manifold transitions. The lines labelled a,b,c,d correspond to transitions between the first and the second manifold, at energies $E_{2d,1d}$, $E_{2u,1d}$, $E_{2d,1u}$, $E_{2u,1u}$. 
Their differences, labelled a-c, b-d, a-b and c-d, coincide two by two. They are shown with red lines and perfectly match the observed low-frequency transitions. The two first ones correspond to a transition energy $E_{1u,1d}$, the two last ones to $E_{2d,2u}$. The lines are dimmer at low frequency because the matrix elements go to zero at phases 0 and $\pi$ (see Fig.~\ref{Fig:Fig7}(d)), and because the difference in occupancy of the two spin states diminishes when their energy difference is comparable to temperature: $k_BT/h\approx 0.8$~GHz. Note that such transitions have been recently driven indirectly through Raman processes \cite{Hays2020}. 

In general, a complete fit of the spectra found in nanowire weak links is not possible with a simple modelling of the weak link. However, in the absence of a drive, the frequency shift of the resonator is often dominated by the contribution from a single channel, which allows for a simpler description.
In Fig.~\ref{Fig:Fig11}(a) and (b), we show the two-tone spectroscopy and single-tone measurement of the same nanowire weak link as in Figs.~\ref{Fig:newspec}, again from another cooldown.
Among the observed transition lines, there is a high-contrast pair transition that crosses the resonator at $\varphi=\pi (1\pm0.12)$.
Within the scattering model (Appendix~\ref{app:scattering}), it can be fitted as a pair transition towards the lowest of three Andreev manifolds arising from a high-transmission channel ($\tau=0.996$). In Fig.~\ref{Fig:Fig11}(a) we indicate this fit with a red line on the right hand side (parameters in Appendix~\ref{app:spiderfits}), as well as two replicas shifted by $\pm f_r$, also visible in the data and associated to a strong measurement tone.  
The corresponding shift of the resonator, fitted with Eq.~(\ref{fShift-Single}) and (\ref{fShift-Ground}) with $\delta_{zp} = 0.012$ and using $f_r=6.60762~$GHz (bare frequency measured when the nanowire is fully depleted), is shown in dashed red in the right hand side of Fig.~\ref{Fig:Fig11}(b). Although it contains contributions from both the continuum and the three Andreev manifolds, it is mainly dominated by the ${\cal H}'_0$ contribution at energy $E_{-1,1}$ associated to the transition to the lowest manifold. Therefore it can be well approximated within a simplified Jaynes-Cummings description, taking Eq.~(\ref{JC}) with a renormalized gap $\Delta_{\text{eff}}/h=15.4~$GHz and $f_A=2E_{-1,1}$ (not shown in the figure for clarity, because it coincides almost exactly with the full theory for the channel shown in dashed red). Although it does not fit perfectly, it offers a simple analytical form that captures well the main features of the data around the anticrossing. The small discrepancies with the experimental data are attributed to the other channels.
Many other transitions are indeed visible in the two-tone spectroscopy, which we model with an effective second channel (parameters in Appendix~\ref{app:spiderfits}). Its contribution, shown with a dashed blue line in Fig.~\ref{Fig:Fig11}(b), produces the smooth background that, added to the shift from the main transition, accounts precisely for the data for all phases (black line).

\section{Conclusions}
We have developed a general formalism to describe the readout of the states of a phase-biased superconducting weak link coupled to a microwave resonator in a circuit-QED setup, based on Ref.~\cite{Park2020}. The fermionic character of the weak link excitations gives rise to a rich phenomenology in the presence of a driving field, with different responses depending on the states' parity. We show how both spin-conserving or non-conserving transitions can occur, depending on how the system is driven. 

We have applied this theory to analyze experimental results obtained on atomic contacts and semiconducting nanowire junctions. For the former, only one pair of spin-degenerate Andreev levels exists per channel, which greatly simplifies the theory, leading to analytical results. Quantitative agreement with the data, both for the ground state properties and the spectroscopy, is reached. For the nanowire case, the junctions host many Andreev states in each channel and spin-degeneracy is broken by spin-orbit coupling. This requires a more involved analysis and the comparison with experiments is less straightforward. However, with a simple modeling of these junctions we were able to interpret several features observed in the experiment, such as the effect on the resonator shift of the levels' curvature and the crossing of virtual transitions with the resonator frequency. In addition, we reported data on longer nanowire weak links with a stronger coupling to a resonator, in which direct intra-manifold spin-flip transitions of a single quasiparticle in a superconducting weak link have been observed in the absence of a Zeeman field. The spectra also suggest that Andreev levels in different channels are coupled, and display groups of similar transitions that require further modelling. 

\appendix

\section{Theory for the frequency shift of a resonator coupled to a driven weak link} \label{app:Spectr}

We provide a detailed derivation of Eqs. \eqref{SQPT} and \eqref{PT} in Sec. \ref{driven-shift}.

\subsection{Dissipation in the resonator and the weak link}
\label{appsub:Diss}
For more realistic description, we assume that external baths are coupled to the system, leading to photon losses in the resonator and relaxation in the weak link. For the resonator, the time evolution of an operator ${\cal O}$ including the photon loss can be described by the master equation 
\begin{equation}
\frac{d {\cal O}}{dt}= i \omega_r[a^{\dagger} a, {\cal O}] 
+ \frac{\kappa (n_t+1)}{2} {\cal L}[a^{\dagger}]{\cal O}+ \frac{\kappa n_t}{2} {\cal L}[a]{\cal O}, 
\end{equation}
where ${\cal L}[c]{\cal O}=2 c {\cal O} c^{\dagger} -c c^{\dagger} {\cal O} - {\cal O} c c^{\dagger}$, $\omega_r = 2 \pi f_r$, $\kappa$ is the loss rate and $n_t = 1/[\text{exp}(\hbar \omega_r/k_B T)-1]$ is the thermal photon in the resonator. Solving the equation for ${\cal O} = a, a^{\dagger}$ and $n_R = a^{\dagger} a$, we obtain
\begin{align}
\begin{split}
a(t) &= e^{-i \omega_r t -\kappa t/2} a(0), \\
a^{\dagger}(t) &= e^{i \omega_r t -\kappa t/2} a^{\dagger}(0), \\
n_R(t) &= e^{-\kappa t} n_R(0) + (1-e^{-\kappa t}) n_t. 
\end{split} \label{Resonator1}
\end{align} 

For the weak link the Hamiltonian from Eq.~(\ref{WLH}) can be written in terms of Bogoliubov operators as 
\begin{equation}
\hat{H}_0= \sum_{i\sigma > 0} E^{}_{i\sigma} \gamma^{\dagger}_{i\sigma} \gamma^{}_{i\sigma} + E_g, 
\end{equation}
where $E_g=(1/2) \sum_{j\sigma'<0} E_{j\sigma'}$ is the ground state energy. In this representation, in order to consider only states with positive energy we used the relation $\gamma^{\dagger}_{i\sigma} = -s \gamma_{-i\bar{\sigma}}$, where $s= 1(-1)$ for $\sigma=u (d)$. The master equation taking into account the relaxation of a quasiparticle from one energy level to the next lower one is given by 
\begin{equation}
\frac{d {\cal O}}{dt}= i \sum_{i\sigma} \omega_{i\sigma}[\gamma^{\dagger}_{i\sigma} \gamma^{}_{i\sigma}, {\cal O}] + \sum_{j\sigma'} \frac{\Gamma_{j\sigma'}}{2}{\cal L}[L_{j\sigma'}]{\cal O},
\end{equation}    
where $\omega_{i\sigma}=E_{i\sigma}/\hbar$, $\Gamma_{j\sigma'}$ is the relaxation rate and  $L_{j\sigma'}=\gamma^{\dagger}_{j\sigma'} \gamma^{}_{j-1 \sigma'}$. If we truncate the equations at first order in $\Gamma_{j\sigma'}$, we obtain approximate solutions, 
\begin{align}
\begin{split}
\gamma_{i\sigma}(t) &= e^{-i \omega_{i\sigma}t - \tilde{\Gamma}_{i\sigma}t/2}\gamma_{i\sigma}(0), \\
\gamma^{\dagger}_{i\sigma}(t) &= e^{i \omega_{i\sigma}t - \tilde{\Gamma}_{i\sigma}t/2}\gamma^{\dagger}_{i\sigma}(0),
\end{split} \label{Weaklink1}
\end{align}
where 
\begin{equation}
\tilde{\Gamma}_{i\sigma} = \Gamma_{i\sigma} (1 -  n_{i-1\sigma}(0)) + \Gamma_{i+1\sigma} n_{i+1 \sigma}(0),
\end{equation}
and $n_{i\sigma}(0)$ is the 
occupancy of the $i\sigma$ Andreev level at $t=0$. For simplicity, we assume that $n_{i-1\sigma}(0)$ and $n_{i+1\sigma}(0)$ are zero, that is, $\tilde{\Gamma}_{i\sigma}= \Gamma_{i\sigma}$.

\subsection{Spectral function of the resonator} \label{appsub:Spectr}

The shift of the resonator coupled to the driven weak link in a many-body state can be obtained by using a retarded Green's function of the resonator, 
\begin{equation}
D^R(\omega) = - i \int^{\infty}_{0} dt e^{i \omega t} \langle \left[ a(t), a^{\dagger}(0)\right]\rangle.  \label{RGF}
\end{equation}
In the absence of the coupling to the weak link, the Green's function is given by $D^R(\omega)=1/(\omega-\omega_r+i \kappa/2)$ where we used the form of $a(t)$ in Eq.~\eqref{Resonator1}. Let us now consider the resonator-weak link coupling Hamiltonian \cite{Park2020} including the driving field, 
\begin{align}
\hat{P}(t)=& \hat{H}_c(\delta) + \hat{A}(t), \\
\hat{H}_c(\delta) =& \hat{\delta}_r \hat{H}'_0(\delta) + \frac{\hat{\delta}^2_r}{2}\hat{H}''_0(\delta),
\end{align}  
where 
$\hat{H}^{\prime (\prime\prime)}_0(\delta)$ is given by 
\begin{align}
\hat{H}^{\prime (\prime\prime)}_0 (\delta)=&
\frac{1}{2} \sum_{i\sigma,j\sigma'} h^{\prime (\prime\prime)}_{i\sigma,j\sigma'} \gamma^{\dagger}_{i\sigma} \gamma_{j\sigma'} \\ 
=& 
\frac{1}{2}\sum_{i\sigma<0} h^{\prime (\prime\prime)}_{i\sigma,i\sigma}+
\sum_{i\sigma,j\sigma'>0} h^{\prime (\prime\prime)}_{i\sigma,j\sigma'} \gamma^{\dagger}_{i\sigma} \gamma_{j\sigma'} \nonumber\\
&- \sum_{i\sigma,j\sigma'>0}\frac{s'}{2}\left( 
h^{\prime (\prime\prime)}_{i\sigma,-j\bar{\sigma}'} \gamma^{\dagger}_{i\sigma} \gamma^{\dagger}_{j\sigma'} + \text{h.c.}\right), \nonumber
\end{align}
and the driving Hamiltonian $\hat{A}(t)$ is  
\begin{align}
\hat{A}(t) &= \frac{1}{2} \sum_{i\sigma < j\sigma'} A_{i\sigma, j\sigma'} \gamma^{\dagger}_{i \sigma} \gamma_{j \sigma'} e^{i \omega_d t} + \text{h.c.} \\
&= \sum_{j\sigma'>i\sigma>0} A_{i\sigma, j\sigma'} \gamma^{\dagger}_{i \sigma} \gamma^{}_{j \sigma'} e^{i \omega_d t} \nonumber\\
&+ \frac{1}{2} \sum_{i\sigma,j\sigma'>0} (-s)A_{-i\bar{\sigma}, j\sigma'} \gamma_{i \sigma} \gamma_{j \sigma'} e^{i \omega_d t} + \text{h.c.}. \nonumber  
\end{align}

In the interaction picture, the expectation value in Eq.~\eqref{RGF} is written as 
\begin{equation}
\langle \left[ a(t), a^{\dagger}(0)\right]\rangle = 
\text{Tr}\left[\rho \left[ \hat{U}^{\dagger}_{\text{I}}(t) a^{}_{\text{I}}(t) \hat{U}^{}_{\text{I}}(t), a^{\dagger}_{\text{I}}(0)\right] \right],\label{Rpropagator}
\end{equation}
where the subscript `I' denotes the interaction picture where creation and annhilation operators evolve in time according to Eqs.~\eqref{Resonator1} and \eqref{Weaklink1}, and $\rho=\rho_R\otimes \rho_{\text{WL}}$ is the density matrix at time $t=0$ with $\rho_R$ for the resonator and $\rho_{\text{WL}}$ for the weak link, where 
\begin{align}
\rho_R&= \sum^{\infty}_{n=0} \frac{e^{-N} N^{n}}{n!} |n \rangle \langle n|,\\
\rho_{\text{WL}}&=  |g\rangle \langle g| ~~ \text{or} ~~ |i_0\sigma_0\rangle \langle i_0\sigma_0|.
\end{align}
Here, $N$ is the mean photon number of the resonator and the many-body state of the weak link at $t=0$ is either the ground ($|g\rangle$) or the excited odd ($|i_0\sigma_0\rangle = \gamma^{\dagger}_{i_0\sigma_0} |g\rangle$) state. The time evolution operator $\hat{U}_{\text{I}}$ in Eq.~\eqref{Rpropagator} is 
\begin{equation}
\hat{U}_{\text{I}}(t) = {\cal T}\left[ \text{exp}\left(\frac{-i}{\hbar} \int^{t}_{0}dt' \; \hat{P}_{\text{I}}(t')\right)  \right],   
\end{equation}
where ${\cal T}$ denotes the time-ordered product. We approximate the evolution operator to obtain $\delta^{2}_{\text{zp}}$-order term (the shift without the driving field) and $\delta^{2}_{\text{zp}}|A_{i\sigma,j\sigma'}|^2$-order term (the shift with the driving field), requiring to expand up to fourth order in $\hat{P}_{\text{I}}(t')$. We assume that the coupling coefficient $\delta_{\text{zp}}$, the strength of the driving field and the dissipation in each system are sufficiently weak such that a leading order correction to the expectation value of \eqref{Rpropagator} can be written as, 
\begin{equation}
\begin{split}
\langle \left[ a(t), a^{\dagger}(0)\right]\rangle &\approx  e^{-i \omega_r t - \kappa t/2} (1-i \delta \omega_r \, t) \\
&\approx e^{-i \omega_r t - i \delta \omega_r t - \kappa t/2}.
\end{split} \label{Correlator}
\end{equation}
Below, we calculate this shifted frequency $\delta \omega_r$, and provide the results given in the main text. Hereafter, all operators will be in the interaction picture (described by  Eqs.~\eqref{Resonator1} and \eqref{Weaklink1}) and we omit the subscript `I' for simplicity unless stated otherwise.  

The perturbation expansion of $\hat{U}^{\dagger}(t) a(t) \hat{U}(t)$ and keeping terms that contribute to the frequency shfit lead to  
\begin{equation}
\begin{split}
\hat{U}^{\dagger}(t) a(t) \hat{U}(t)=& a(t) \big[ 1+ C_{H''}(t) + C_{H'^2}(t) \\
&+ C_{H''A^2}(t) +C_{H'^2A^2}(t)\big], 
\end{split} \label{U-expansion}
\end{equation}
where $C_{H''}(t)$ and $C_{H'^2}(t)$ are $\delta^{2}_{\text{zp}}$-order terms,
\begin{align}
C_{H''}(t) =&\frac{-i \delta^2_{\text{zp}}}{\hbar} \int^{t}_{0} dt_1 \, H''(t_1),\\
C_{H'^2}(t)=&\frac{\delta^2_{\text{zp}}}{\hbar^2} \int^{t}_{0} dt_1 \int^{t}_{0} dt_2 
\left(-\theta_{12} e^{i W_{12}} +\theta_{21} e^{i W_{21}} \right) \nonumber\\
& \times H'(t_1) H'(t_2),
\end{align}
where $\theta_{ij}=\theta(t_i-t_j)$ is the Heaviside step-function and
\begin{align}
W_{ij}= (\omega_r + i \kappa/2) t_i+(- \omega_r +i \kappa/2) t_j.
\end{align}
The field-dependent terms $C_{H''A^2}(t)$ and $C_{H'^2A^2}(t)$ are given by 
\begin{widetext}
\begin{align}
C_{H''A^2}(t)=& \frac{i \delta^2_{\text{zp}}}{\hbar^3} 
\int^{t}_{0} dt_1 \int^{t}_{0} dt_2 \int^{t}_{0} dt_3 
\,\theta_{12}\theta_{23} \, H''_1 A_2 A_3 
-\theta_{21}\theta_{23} \, A_1 H''_2 A_3 
+\theta_{21}\theta_{32} \, A_1 A_2 H''_3, \\
C_{H'^2A^2}(t)=& \frac{\delta^2_{\text{zp}}}{\hbar^4} 
\int^{t}_{0} dt_1 \int^{t}_{0} dt_2 \int^{t}_{0} dt_3 \int^{t}_{0} dt_4 \nonumber\\
& \theta_{12}\theta_{23}\theta_{34} \left( e^{i W_{12}}  H'_1 H'_2 A_3 A_4
+ e^{i W_{13}}  H'_1 A_2 H'_3 A_4
+ e^{i W_{14}}  H'_1 A_2 A_3 H'_4\right)\nonumber\\
& -\theta_{21}\theta_{23}\theta_{34} \left( e^{i W_{21}} H'_1 H'_2 A_3 A_4
 + e^{i W_{23}} A_1 H'_2 H'_3 A_4
+ e^{i W_{24}} A_1 H'_2 A_3 H'_4\right)\nonumber\\
& +\theta_{21}\theta_{32}\theta_{34} \left( e^{i W_{31}} H'_1 A_2 H'_3 A_4
 + e^{i W_{32}} A_1 H'_2 H'_3 A_4
+e^{i W_{34}} A_1 A_2 H'_3 H'_4 \right)\nonumber\\
& -\theta_{21}\theta_{32}\theta_{43} \left( e^{i W_{41}} H'_1 A_2 A_3 H'_4
+e^{i W_{42}} A_1 H'_2 A_3 H'_4
+e^{i W_{43}} A_1 A_2 H'_3 H'_4\right),
\end{align}
\end{widetext} 
where the subscript numbers $n=1,...4$ indicate the time dependence, 
\begin{align}
H''_n = H''(t_n), \,\, H'_n = H'(t_n), \,\, A_n = A(t_n).
\end{align}
Substituting Eq.~\eqref{U-expansion} into Eq.~\eqref{Correlator} and performing the integrations with respect to the given many-body state, we find the shifted frequency $\delta \omega_r$, 
\begin{align}
\delta \omega_r = \delta \omega_{H''}+\delta \omega_{H'^2}
+\delta \omega_{H''A^2}+ \delta \omega_{H'^2A^2}, 
\end{align}
where $\delta\omega_X$, with $X \in \left\{H'',H'^2,H''A^2,H'^2A^2 \right\}$, are the frequency assciated with $C_{X}(t)$, 
\begin{align}
\delta\omega_X = \frac{i}{t} \text{Tr}\left[\rho_{\text{WL}} \, C_X(t) \right]. 
\end{align}

For $\rho_{\text{WL}} = |g\rangle \langle g|$, they are given by  
\begin{align}
\delta \omega_{H''}\left. \right|_{|g\rangle}=& \frac{\delta^2_{\text{zp}}}{2 \hbar} \sum_{i\sigma<0} h''_{i\sigma,i\sigma},\\
\delta \omega_{H'^2}\left. \right|_{|g\rangle}=& \frac{\delta^2_{\text{zp}}}{2 \hbar} \sum_{i\sigma<0 } \mathcal{R}_{i\sigma},\\
\delta \omega_{H''A^2}\left. \right|_{|g\rangle}=&\frac{\delta^2_{\text{zp}}}{\hbar} \sum_{i\sigma>0} \sum_{ j\sigma'>0} 
\frac{|A_{-i \bar{\sigma},j\sigma'}|^2}{|D_{-i \bar{\sigma},j\sigma'}|^2} \\
& \times\left(h''_{i \sigma,i \sigma} + h''_{j \sigma',j \sigma'} \right),\nonumber\\
\delta \omega_{H'^2A^2}\left. \right|_{|g\rangle}=& \frac{\delta^2_{\text{zp}}}{\hbar} \sum_{i\sigma>0 } \sum_{j\sigma'>0} 
\frac{|A_{-i \bar{\sigma},j\sigma'}|^2}{|D_{-i \bar{\sigma},j\sigma'}|^2} \\
& \times \left(\mathcal{R}_{i\sigma} + \mathcal{R}_{j\sigma'}
\right),\nonumber
\end{align} 
where
\begin{align}
\mathcal{R}_{i\sigma}&=-\sum_{j\sigma' \neq i\sigma} 
\frac{2 |h'_{i\sigma,j\sigma'}|^2 E_{i\sigma,j\sigma'}}{E^2_{i\sigma,j\sigma'}-\tilde{E}_r^2},\\
D_{i\sigma,j\sigma'}&=\hbar \omega_d -  |E_{j\sigma',i\sigma}|
+ i (\Gamma_{i\sigma}+\Gamma_{j\sigma'}) \hbar/2.
\end{align} 
Here, $E_{j\sigma',i\sigma} = E_{i\sigma}-E_{j\sigma'} $ is the energy difference between the levels $i\sigma$ and $j\sigma'$, and $\tilde{E}_r$ is the resonant energy of the resonator including the effect of the dissipation, $\tilde{E}_r = h f_r - i (\Gamma_{i\sigma}+\Gamma_{j\sigma'}+\kappa)/2$. 

The frequency shift for $\rho_{\text{WL}} = |i_0\sigma_0\rangle \langle i_0 \sigma_0|$ is obtained as 
\begin{align}
\delta \omega_{H''}\left. \right|_{|i_0\sigma_0\rangle}=& 
\delta \omega_{H''}\left. \right|_{|g\rangle }  +\frac{\delta^2_{\text{zp}}}{\hbar}  h''_{i_0\sigma_0,i_0\sigma_0},\\
\delta \omega_{H'^2}\left. \right|_{|i_0\sigma_0\rangle}=& 
\delta \omega_{H'^2}\left. \right|_{|g\rangle}+
\frac{\delta^2_{\text{zp}}}{\hbar} \mathcal{R}_{i_0 \sigma_0},\\
\delta \omega_{H''A^2}\left. \right|_{|i_0\sigma_0\rangle}=&\frac{2\delta^2_{\text{zp}}}{\hbar}
\sum_{j\sigma'>0} 
\frac{|A_{i_0\sigma_0,j\sigma'}|^2}{|D_{i_0 \sigma_0,j\sigma'}|^2} \\
&\times \left(h''_{j \sigma',j \sigma'} - h''_{i_0 \sigma_0,i_0 \sigma_0} \right),\nonumber\\
\delta \omega_{H'^2A^2}\left. \right|_{|i_0\sigma_0\rangle} =& \frac{2 \delta^2_{\text{zp}}}{\hbar} \sum_{j\sigma'>0} 
\frac{|A_{i_0\sigma_0,j\sigma'}|^2}{|D_{i_0 \sigma_0,j\sigma'}|^2} \\ 
&\times\left( 
\mathcal{R}_{j\sigma'}-\mathcal{R}_{i_0 \sigma_0}\nonumber
\right).
\end{align} 

First, note that in absence of drive, the coupling to the weak link induces a shift in the resonator's frequency which is
\begin{eqnarray*}
\left.\frac{\hbar\delta \omega_r^{|g\rangle}}{\delta^2_{\text{zp}}}\right|_{\text{dr off}} &=& \frac{1}{2} \sum_{i\sigma<0} \left( h''_{i\sigma,i\sigma} + \mathcal{R}_{i\sigma} \right) \\
&=&\frac{1}{2} \sum_{i\sigma<0} \left[ h''_{i\sigma,i\sigma} - \sum_{j\sigma' >0} 
\frac{2 |h'_{i\sigma,j\sigma'}|^2 E_{i\sigma,j\sigma'}}{E^2_{i\sigma,j\sigma'}-\tilde{E}_r^2} \right],\\
\end{eqnarray*}
which is a generalized version of Eq.~(\ref{fShift-Ground2}) taking into account the intrinsic resonator's linewidth and the inherited contribution due to the finite life-time of Andreev states. To see this, note that the last term can be written as
\begin{eqnarray*}
- |h'_{i\sigma,j\sigma'}|^2 &&\left( 
\frac{1}{E_{i\sigma,j\sigma'}-h f_r + i (\Gamma_{i\sigma}+\Gamma_{j\sigma'}+\kappa)/2} \right.\\
&+&\left. \frac{1}{E_{i\sigma,j\sigma'}+h f_r - i (\Gamma_{i\sigma}+\Gamma_{j\sigma'}+\kappa)/2} \right),
\end{eqnarray*}
and has the form of Jaynes-Cummings with dissipation.
The contribution from the drive is
\begin{eqnarray*}
\left.\frac{\hbar\delta\omega_r^{|g\rangle}}{\delta^2_{\text{zp}}}\right|_{\text{dr on}} &=& \left.\frac{\hbar\delta\omega_r^{|g\rangle}}{\delta^2_{\text{zp}}}\right|_{\text{dr off}} + \sum_{i\sigma>0} \sum_{ j\sigma'>0} 
\frac{|A_{-i \bar{\sigma},j\sigma'}|^2}{|D_{-i \bar{\sigma},j\sigma'}|^2}\\ &\times& \left(h''_{i \sigma,i \sigma} + \mathcal{R}_{i\sigma} + h''_{j \sigma',j \sigma'} + \mathcal{R}_{j\sigma'} \right),
\end{eqnarray*}
which for example, if the drive frequency is close to $|g\rangle \rightarrow |1u 1d\rangle$ transition, the term with the denominator
\begin{equation*}
D_{-1u,1u}=\hbar \omega_d -  |E_{1u,-1u}|
+ i (\Gamma_{-1u}+\Gamma_{1u}) \hbar/2,
\end{equation*}
becomes the most relevant and
\begin{align*}
&\left.\frac{\hbar\delta\omega_r^{|g\rangle}}{\delta^2_{\text{zp}}}\right|_{\text{dr on}} = \left.\frac{\hbar\delta\omega_r^{|g\rangle}}{\delta^2_{\text{zp}}}\right|_{\text{dr off}} \\
&+\frac{|A_{-1u,1u}|^2\  \left(h''_{1u,1u} + \mathcal{R}_{1u} + h''_{1d,1d} + \mathcal{R}_{1d} \right) }{\left( \hbar \omega_d -  |E_{1u,-1u}|\right)^2+(\Gamma_{-1u}+\Gamma_{1u})^2 \hbar^2/4},
\end{align*}
which corresponds to the shift in $|1u 1d\rangle$ multiplied by a lorenztian coefficient associated with the probability of the transition, here the matrix element is the Rabi frequency. So the result is very intuitive: once the shifts $\left.\delta\omega_r^{|\Psi\rangle}\right|_{\text{dr off}}$ for the different many-body configurations are determined, the effect of the drive is to take into account selection rules for transitions to occur.

\section{Theoretical model of the nanowire: scattering model}
\label{app:scattering}
We briefly repeat the theoretical model of the nanowire weak link reported in Refs.~\cite{Park2017,Tosi2019}, and discuss the effect of side gates on the transitions between Andreev levels of the weak link.

The nanowire is described by the Hamiltonian ${\cal H}^{\text{3D}}$ consisting of a kinetic energy, a confining harmonic potential in the transverse $y$ and $z$-directions with a confinement width $W$ (effective diameter of the nanowire) and Rashba spin-orbit coupling with intensity $\alpha$,  
\begin{equation}
{\cal H}^{\text{3D}} = \frac{p^{2}_x+p^{2}_y+p^{2}_z}{2 m^*} + \frac{ \hbar^2 (y^2 + z^2)}{2 m^* (W/2)^4} + \alpha (-k_x s_y + k_y s_x), 
\label{ModelHamiltonian}
\end{equation} 
where $m^*$ is the effective mass and $s_{x,y}$ are the Pauli matrices for spin. We consider the lowest two spin-full transverse subbands denoted by $ns$, with $n=1,2$ and $s=\uparrow,\downarrow$, arising from the confining potential in the transverse direction under the effect of the Rashba spin-orbit coupling \cite{Park2017}. The transverse mode wave functions $\phi^{\perp}_{ns}(y,z)$ are  
\begin{equation}
\begin{split}
\phi^{\perp}_{1s}(y,z)&= \frac{2}{\sqrt{\pi} W} e^{-2(y^2+z^2)/W^2} \chi_s, \\
\phi^{\perp}_{2s}(y,z)&= \frac{2 \sqrt{2} y}{\sqrt{\pi} W^2} e^{-2(y^2+z^2)/W^2} \chi_s,
\end{split} \label{Tmode}
\end{equation}
with energy eigenvalues $E^{\perp}_n = 4\hbar^2 n/(m^* W^2)$, where $\chi_{\uparrow(\downarrow)}=(1/\sqrt{2}) [1,i(-i)]$. By integrating out the transverse modes, ${\cal H}^{\text{3D}}$ can be reduced to one-dimensional form. The energy dispersion relations of the resulting lowest subbands (named $m_1$ and $m_2$) are  
\begin{equation}
E_{\nu}(k_x) = \frac{\hbar^2 k^2_x}{2 m^*} + \frac{E^{\perp}_{+}}{2} 
- \sqrt{\left(  \frac{E^{\perp}_{-}}{2} +(-1)^{\nu} \alpha k_x\right)^2 + \eta^2}, 
\label{Dispersion}
\end{equation} 
where $\nu=1$ corresponds to $m_1$ and $\nu=2$ to $m_2$, and $E^{\perp}_{\pm}=E^{\perp}_1 \pm E^{\perp}_2$. $\eta=\sqrt{2}\alpha/W$ is the strength of the subband mixing due to the Rashba spin-orbit coupling. In accordance to the estimated nanowire diameter in Ref.~\cite{Tosi2019} we take $W\sim 140$ nm, which leads to $E^{\perp}_2-E^{\perp}_1 \sim 0.68$~meV for the subband separation.

\subsection{Andreev levels}\label{app:ABS}
The linearized Bogoliubov-de Gennes equation around the chemical potential $\mu$ for which only the lowest subbands are occupied is 
\begin{equation}
\begin{gathered}
{\cal H}_0 \Phi(x) = E \,\Phi(x), \\
{\cal H}_0 = \left({\cal H}_{\text{NW}} + {\cal H}_b + {\cal H}_A\right) \hat{\tau}_z + \Delta(x)\hat{\tau}_x, 
\end{gathered}\label{BdG}
\end{equation}
where $\hat{\tau}_{x,z}$ are Pauli matrices in Nambu space. The wave function is written in the basis
\begin{equation}
\Phi=\left(\phi^{e}_{+,R},\phi^{e}_{+,L},\phi^{e}_{-,R},\phi^{e}_{-,L},\phi^{h}_{+,R},
\phi^{h}_{+,L},\phi^{h}_{-,R},\phi^{h}_{-,L}\right)^T,\nonumber
\end{equation}
where $R(L)$ refers to the right-moving (left-moving) electron $(e)$ or hole $(h)$ in the bands $m_1$($-$), $m_2$($+$). The Hamiltonian for electrons in the nanowire ${\cal H}_{\text{NW}}$ and the potential scattering term ${\cal H}_b$ are 
\begin{equation}
\begin{split}
{\cal H}_{\text{NW}}&= -i \left( v_{+} \hat{d}_z + v_{-} \hat{b}_z \right) \hbar \partial_x 
-\hbar (k_{+} + k_{-} \hat{b}_z \hat{d}_z),\\
{\cal H}_b&= U_0 \delta(x-x_0) \left( 1+ \hat{d}_x \text{cos}((\theta_1-\theta_2)/2)\right),
\end{split}
\end{equation}
where $\hat{b}_j$ and $\hat{d}_j$ ($j=x,y,z$) are Pauli matrices acting in the subband ($+,-$) and right/left mover space, respectively. $v_{\pm}$ and $k_{\pm}$ are functions of Fermi velocities ($v_{1}$ and  $v_{2}$) and Fermi wave vectors ($k_{F1}$ and $k_{F2}$), 
\begin{align}
v_{\pm} = \frac{v_1 \pm v_2}{2}, \hspace{10pt}
k_{\pm} = \frac{v_1 k_{F1} \pm v_2 k_{F2}}{2},
\end{align} 
where $\hbar v_j = \partial E_{j}(k_{Fj})/\partial k_x$ with $E_{j}(k_{Fj})=\mu$. Particle backscattering within the nanowire is accounted for by a single delta-like potential barrier of the strength $U_0$ located at some arbitrary position $x_0$, and $\theta_{j=1,2}=\arccos[(-1)^j (\hbar k_{F_j}/m^*-v_j)/\alpha]$ characterize the mixing with the higher subbands.  
The superconducting order parameter $\Delta(x)$ is given by $\Delta(x)=\Delta$ at $|x|>L/2$ and zero otherwise. We assume that the phase drop of the superconducting order parameter across the weak link occurs at $x=x_0$.

The Hamiltonian ${\cal H}_0$ commutes with the pseudospin operator $b_z$ which defines pseudospin-up and -down Andreev states $\Phi_{\sigma}(x) = \tilde{\Phi}_{\sigma}(x)\otimes|\sigma\rangle$ with $\sigma=u, d$, 
\begin{equation}
\begin{split}
\tilde{\Phi}_{u} &=\left(\phi^{e}_{+,R},\phi^{e}_{+,L},\phi^{h}_{+,R},
\phi^{h}_{+,L}\right)^T,\\
\tilde{\Phi}_{d} &=\left(\phi^{e}_{-,R},\phi^{e}_{-,L},\phi^{h}_{-,R},\phi^{h}_{-,L}\right)^T.
\end{split}
\end{equation}  
The Andreev states are obtained by imposing the boundary condition at $x=x_0$, 
\begin{equation}
\tilde{\Phi}_{\sigma}(x_0+0^{+}) = 
\hat{M}_{\sigma}
\tilde{\Phi}_{\sigma}(x_0-0^{+}),
\label{BDx0}
\end{equation}
where $0^{+}$ is a positive infinitesimal. $\hat{M}_{\sigma}$ is the transfer matrix given by 
\begin{equation}
\hat{M}_{u} = 
\begin{pmatrix}
e^{- i \delta/2} M_{12} & 0  \\
0 & e^{i \delta/2} M_{12} \\
\end{pmatrix},
\end{equation}
and $\hat{M}_{d}$ by replacing $M_{12}$ in $\hat{M}_{u}$ by $M_{21}$, where $M_{ij}$ is the $2 \times 2$ matrix given by
\begin{equation}
M_{ij} = \dfrac{1}{t'}
\begin{pmatrix}
t t'-r r'& \sqrt{\dfrac{v_j}{v_i}} r' e^{i \varphi} \\
-\sqrt{\dfrac{v_i}{v_j}} r  e^{-i \varphi}&  1
\end{pmatrix}
\end{equation} 
with $\varphi=((k_{F1}+ k_{F2})+(\lambda_1+\lambda_2)\epsilon/L) x_0.$
The reflection and transmission coefficients are determined by  
\begin{align}
t e^{-i u_a}  &= t' e^{i u_a}  = \left(\cos d + i u_s\dfrac{\sin d}{d} \right)^{-1}, \nonumber\\
 r e^{-i \varphi} &=  r' e^{i \varphi}= -i  \sqrt{u_1 u_2}\frac{\sin d}{d} 
\cos \left(\frac{\theta_1-\theta_2}{2}\right) \sqrt{t t'}, \nonumber\\
d &= \frac{1}{2}\sqrt{u^2_1+u^2_2 -  2 u_1 u_2 \cos(\theta_1-\theta_2)},
\label{TandR}
\end{align}
where $v_0 = \hbar v_1 v_2/U_0$, $u_j=v_j/v_0,$ $u_s = (u_1+u_2)/2,$ and $u_a = (u_1-u_2)/2$. From the continuity conditions at $x=\pm L/2$ and Eq.~\eqref{BDx0} we find the transcendental equation
\begin{eqnarray}
\tau\cos\left((\lambda_{1}-\lambda_{2})\epsilon \mp\delta\right) + (1-\tau) \cos((\lambda_{1}+\lambda_{2}) \epsilon x_r) &=& \nonumber \\
 \cos(2 \arccos \epsilon - (\lambda_{1}+\lambda_{2})\epsilon), ~~
&& \;
\label{eq:transcendental}
\end{eqnarray} 
where $x_r = 2x_0/L$ and $\tau = |t|^2$ is the transmission probability. The normalization condition for the Andreev bound states $\Phi_{i\sigma}$ with $|E_{i\sigma}|<\Delta$ is 
\begin{equation}
\int^{\infty}_{-\infty} dx \, \Phi^{\dagger}_{i \sigma}(x) \Phi_{j \sigma'}(x) = \delta_{ij} \delta_{\sigma \sigma'}.
\end{equation} 
The continuum states $\Phi_{i\sigma}$ at $|E|\geq \Delta$ are degenerate with $i=(i_p,i_d)$ where $i_p=e, h$ denote electron-like or hole-like state and $i_d=l,r$ are the left or right source. They are normalized by the local density at the source region, 
\begin{equation}
\Phi^{\dagger}_{i\sigma}(x) \Phi_{j\sigma'}(x) = \frac{\delta_{ij} \delta_{\sigma \sigma'}}{2 \pi \hbar v_g(E)},
\end{equation}
where the group velocity $v_g(E)$ at energy $E$ is given by $v_1 \sqrt{E^2-\Delta^2}/|E|$ for $(\sigma,i_p,i_d)=(u,e,l),$ $(u,h,r),$ $(d,e,r),$ $(d,h,l)$, and $v_2 \sqrt{E^2-\Delta^2}/|E|$ for $(\sigma,i_p,i_d)=(u,e,r),$ $(u,h,l),$ $(d,e,l),$ $(d,h,r)$.

The current operator ${\cal H}'_0=\partial {\cal H}_0 /\partial \delta$ of the linearized band is given by
\begin{equation}
{\cal H}'_0 = \frac{\hbar \delta(x-x_0)}{2}
\begin{pmatrix}
 v1 & 0 & 0 & 0 \\
  0 & -v2 & 0 & 0 \\
  0 & 0 & v2 & 0 \\
  0 & 0 & 0 & -v1 
\end{pmatrix} 
 \hat{\tau}_0,
\end{equation}
where the factor $1/2$ reflects the particle hole redundancy and $\hat{\tau}_0$ is the identity matrix in the Nambu space. We find the matrix element of the current operator between states $i\sigma$ and $j\sigma'$. For pseudospin-up states, $\sigma, \sigma' = u$, we obtain
\begin{eqnarray}
\langle \Phi_{iu}| {\cal H}'_0|\Phi_{ju}\rangle &=& 
\int \,dx\, \Phi^{\dagger}_{iu}(x) \,{\cal H}'_0 \,\Phi_{ju}(x) \\
&=& \frac{\hbar}{2} \tilde{\Phi}^{\dagger}_{iu}(x_0) 
\begin{pmatrix}
v1 & 0 & 0 & 0 \\
  0 & -v2 & 0 & 0 \\
  0 & 0 & v1 & 0 \\
  0 & 0 & 0 & -v2
\end{pmatrix}
\tilde{\Phi}_{ju}(x_0),\nonumber
\end{eqnarray}
and for pseudospin-down sates, $\sigma, \sigma' = d$, 
\begin{eqnarray}
\langle \Phi_{id}| {\cal H}'_0|\Phi_{jd}\rangle &=& 
\int \,dx\, \Phi^{\dagger}_{id}(x) \,{\cal H}'_0 \,\Phi_{jd}(x) \\
&=& \frac{\hbar}{2} \tilde{\Phi}^{\dagger}_{id}(x_0) 
\begin{pmatrix}
v2 & 0 & 0 & 0 \\
  0 & -v1 & 0 & 0 \\
  0 & 0 & v2 & 0 \\
  0 & 0 & 0 & -v1
\end{pmatrix}
\tilde{\Phi}_{jd}(x_0).\nonumber
\end{eqnarray}

\subsection{Effect of side gates and selection rules}\label{app:Gates}

A displacement $\delta V(\vec{r})$ in the electrostatic potential of the nanowire induced by the side gates shown in Fig. \ref{Fig:Fig7} can be modeled as 
\begin{equation}
\delta V(\vec{r})=
\begin{cases}
\delta V(y) \,\,\,\,&\text{for $x_1 < x < x_2$},\\
0                &\text{elsewhere} , 
\end{cases}
\end{equation}
where we assumed that the electric field of the gates is in the $y$-direction and are uniform in the normal region $x_1 < x < x_2$. To reveal the role of the spatial symmetry of the transverse wave functions, we decompose $\delta V(y)$ into symmetric and anti-symmetric parts, $\delta V(y) = \delta V_{\text{S}}(y) + \delta V_\text{A}(y)$, 
\begin{equation}
\begin{split}
\delta V_{\text{S}}(y) &= \delta V_{\text{S}}(-y) = \frac{\delta V(y) + \delta V(-y)}{2}, \\
\delta V_{\text{A}}(y) &= - \delta V_{\text{A}}(-y)=\frac{\delta V(y) - \delta V(-y)}{2}. 
\end{split}\label{VSymmetry}
\end{equation}
Mapping onto the subspace spanned by the transverse modes, $(\phi^{\perp}_{1s}, \phi^{\perp}_{2s})$, given in Eq.~\eqref{Tmode}, leads to  
\begin{align}
\begin{split}
&\int dydz \,\phi^{\perp \dagger}_{1s}(y,z) \, \delta V_{\text{S}}(y) \,\phi^{\perp}_{1s'}(y,z) = V_{\text{S1}}\, \delta_{ss'}, \\
&\int dydz \,\phi^{\perp \dagger}_{2s}(y,z) \, \delta V_{\text{S}}(y) \,\phi^{\perp}_{2s'}(y,z) = V_{\text{S2}} \,\delta_{ss'}, \\
&\int dydz \,\phi^{\perp \dagger}_{1s}(y,z) \, \delta V_{\text{A}}(y) \,\phi^{\perp}_{2s'}(y,z) = V_{\text{A}} \,\delta_{ss'},  \\
&\int dydz \,\phi^{\perp \dagger}_{1s}(y,z) \, \delta V_{\text{A}}(y) \,\phi^{\perp}_{1s'}(y,z) = 0, \\
&\int dydz \,\phi^{\perp \dagger}_{2s}(y,z) \, \delta V_{\text{A}}(y) \,\phi^{\perp}_{2s'}(y,z) = 0, \\
&\int dydz \,\phi^{\perp \dagger}_{1s}(y,z) \, \delta V_{\text{S}}(y) \,\phi^{\perp}_{2s'}(y,z) = 0, 
\end{split}
\end{align}
where we used the spatial symmetry of the transverse modes in the $y$ direction. Here, $V_{\text{S1}}$ and $V_{\text{S2}}$ are associated with the symmetric part $\delta V_{\text{S}}(y)$, and, $V_{\text{A}}$ on the anti-symmetric part $\delta V_{\text{A}}(y)$, with which we fit the experimental data. ${\cal H}_A$ is given by
\begin{equation}
{\cal H}_A = r(x)
\left(A_{\text{S1}} + A_{\text{S2}} \hat{b}_z \hat{d}_z + A_{\text{S3}} \hat{d}_x
+ A_{\text{A}} \hat{b}_x \hat{d}_z\right),\label{Hgate}
\end{equation}
where $r(x)$ is the rectangular function that takes $1$ for $x_1 < x < x_2$ and $0$ elsewhere, and  
\begin{align}
A_{\text{S1}} &= \sum_{n=1,2} \frac{V_{\text{S1}}}{2}\sin^2 \frac{\theta_n}{2}+ \frac{V_{\text{S2}}}{2} \cos^2 \frac{\theta_n}{2},\nonumber\\
A_{\text{S2}} &= \sum_{n=1,2} (-1)^{n+1} \left(\frac{V_{\text{S1}}}{2}\sin^2 \frac{\theta_n}{2}+ \frac{V_{\text{S2}}}{2} \cos^2 \frac{\theta_n}{2}\right),\nonumber\\
A_{\text{S3}} &= V_{\text{S1}} \sin\frac{\theta_1}{2} \sin\frac{\theta_2}{2} + V_{\text{S2}} \cos\frac{\theta_1}{2}\cos\frac{\theta_2}{2}, \nonumber\\
A_{\text{A}} &= V_{\text{A}} \sin \left(\frac{\theta_1-\theta_2}{2}\right).\nonumber
\end{align}
Note that the last term in Eq.~\eqref{Hgate} containing the matrix $\hat{b}_x$ does not conserve the pseudospin $\sigma$, and thus, are responsible for the pseudospin-flip transitions. Specifically, 
the matrix elements for the pseudospin conserving transitions take the forms
\begin{align}
A_{iu,ju}&=\int^{x_2}_{x_1} dx \,
\tilde{\Phi}_{iu}^{\dagger}(x) \hat{A}_{\text{S}+} \hat{\tau}_z
\tilde{\Phi}_{ju}(x),\\
A_{id,jd}&=\int^{x_2}_{x_1} dx \,
\tilde{\Phi}_{id}^{\dagger}(x) 
\hat{A}_{\text{S}-} \hat{\tau}_z
\tilde{\Phi}_{jd}(x),
\end{align} 

where $\hat{A}_{\text{S}\pm}=A_{\text{S1}}\pm A_{\text{S2}} \hat{d}_z + A_{\text{S3}} \hat{d}_x$.
The matrix element between Andreev states of opposite pseudospin is 
\begin{align}
A_{iu,jd}= \int^{x_2}_{x_1} dx \, 
\tilde{\Phi}_{iu}^{\dagger}(x) A_{\text{A}} \hat{d}_z \hat{\tau}_z \tilde{\Phi}_{jd}(x),
\end{align}   
indicating that the anti-symmetric parameter $V_{\text{A}}$ allows the pseudospin flip transitions. For the matrix elements shown in Figs.~\ref{Fig:Fig7}(d) and \ref{Fig:FitPRX2}, we used the parameters $(V_{\text{S1}},V_{\text{S2}},V_{\text{A}})$ which are $(324\,\mu\text{eV},130\,\mu\text{eV},0)$ and $(0,0,324\,\mu\text{eV})$ for the symmetric and anti-symmetric cases, respectively, and $(10\,\mu\text{eV},4\,\mu\text{eV},314\,\mu\text{eV})$ for the almost anti-symmetric case. $x_1=-108\,\text{nm}$ and $x_2=166\,\text{nm}$ are taken for all cases. For Fig.~\ref{Fig:FitPRX1}, we used $(V_{\text{S1}},V_{\text{S2}},V_{\text{A}})=(10.8\,\mu\text{eV},-7.2\,\mu\text{eV},216.2\,\mu\text{eV})$, $x_1=-10.3\,\text{nm}$ and $x_2=181\,\text{nm}$. 

\section{Tight-binding modelling}
\label{app:TB}
The tight-binding calculations used in Fig.~\ref{Fig:FigLongJunction} use customary Nambu-spinor discretization, in particular the one presented in Ref.~[\onlinecite{Reynoso:PRB:2013}]. As the goal is a qualitative understanding of the phenomena we model the effect of spin-orbit channel mixing by considering only the two transverse channels of a two-sites-wide tight-binding stripe. The longitudinal hopping is taken $t_\parallel= 35.3\Delta$ and the perpendicular hopping $t_\perp=0.3 t_\parallel$. The longitudinal (perpendicular) spin-orbit coupling, which is only included in the normal region, involves $\sigma_y$ ($\sigma_x$) and has a hopping strength of $\lambda_\parallel=0.43 t_\parallel$ ($\lambda_\perp=\sqrt{0.3}\lambda_\parallel$). The chemical potential lies at $0.573 t_\parallel$ from the bottom of the band. It has been tuned to allow the transport through a single effective channel. This choice also ensures a significant difference in the spin-dependent Fermi velocities \cite{Reynoso2012}. The normal region is modeled with 80 sites, whereas 350 sites represent each superconducting region. A direct calculation of the Fermi velocity allows estimating $L/\xi_N\approx 1.6$. Finally, the superconducting phase difference $\delta$ enters (halved) in the longitudinal hopping-terms at the center of the normal region \cite{HamiltonianPRB1996}, and a sharp uncentered barrier of height $\frac{2}{3} t_\parallel$ is located at site 20 of the normal region.

\section{Link between applied flux and phase across the weak link}
\label{app:screening}
When a weak link is placed inside a superconducting loop with geometric inductance $\ell$ threaded by a magnetic flux $\Phi$, the screening current due to the weak link leads to a phase drop across the loop inductance. This leads to the following relation between the reduced flux $\varphi=2\pi \Phi/\Phi_0$ and the phase $\delta$ across the weak link: 
\begin{equation}
    \delta=\varphi-\beta i_{|\Psi\rangle}(\delta),
\end{equation}
with the screening parameter $\beta=\ell \Delta / \varphi_0^2$, $\varphi_0=\Phi_0/2\pi,$ and $i_{|\Psi\rangle}=\varphi_0 I_{|\Psi\rangle}/\Delta=(1/\Delta)\partial E_{|\Psi\rangle}/\partial \delta$ the reduced current associated to the weak link in state $|\Psi\rangle$.
In a single-channel short weak link, $i_{|g\rangle}=-\partial \sqrt{1-\tau \sin^2(\delta/2)}/\partial \delta$, $i_{|e\rangle}=-i_{|g\rangle}$ and $i_{|o\rangle}=0$.

\section{Parameters for Fig.~\ref{Fig:fitAC}}
The parameters used for Fig.~\ref{Fig:fitAC} are given in Table~\ref{Table:Transmissions}.
\label{app:Transmissions}

\begin{table}[h!]
\begin{center}
\begin{tabular}{ |c | c c c c c c| } 
 \hline
  contact & run & $f_r$~(GHz) & $\tau_1$ & $\tau_2$ & $\tau_3$ &$f_{A1}^{\rm min}$~(GHz)\\
 \hline
 (a) & I & 10.1345 & 0.9850	& 0.9428	& 0.9428 & 10.85\\ 
 (b) & I &	10.1345	& 0.9856	& 0.9468	& 0.9468 & 10.63\\
 (c) & III &	10.1091	& 0.9890	& 0.9	& - & 9.29\\
 (d) & I &	10.1345	& 0.9922	& 0.8783	& - & 7.82\\
 (e) & III &	10.1091	& 0.9945	& 0.6561	& - & 6.57\\
 (f) & II &	10.1364& 	0.9967	& 0.9692	& - & 5.09\\
 (g) & I &	10.1345	& 0.9996	& 0.8497	& - & 1.77\\
 \hline
 \end{tabular}
\end{center}
\caption{The data of Fig.~\ref{Fig:fitAC} were taken during different cooldowns (runs) of the same sample, labelled I, II and III. In runs I and III, the bare resonator frequency $f_r$ was measured when the contact was open, whereas for run II it was a fit parameter. The largest transmission $\tau_1$ is the essential fit parameter, determining the overall shape. In the last column, we indicate the minimal value of the Andreev frequency associated to this transmission $f_{A1}^{\rm min}=f_{A1}(\pi)=2\Delta \sqrt{1-\tau_1}.$ When $f_{A1}^{\rm min}<f_r$ (c-g), one observes an avoided crossing. A second channel with transmission $\tau_2$, and for Fig.~\ref{Fig:fitAC}(a',b') a third one  with transmission $\tau_3$ taken equal to $\tau_2$, essentially accounts for an overall shift of the resonator frequency. From the estimated loop inductance $\ell\approx 0.1~$nH \cite{thesisJanvier}, we obtain the screening parameter $\beta=0.03$ used to fit the data.}
\label{Table:Transmissions}
\end{table}

\section{Spectra parameters}
\label{app:spiderfits}
Most of the spectra (illustrative of fit) were obtained from Eq.~(\ref{eq:transcendental}). We give in Table~\ref{Table:spiderfit} the parameters corresponding to the various figures.
\begin{table}[h!]
\begin{center}
\begin{tabular}{ |c | c c c c| } 
 \hline
  ~ & $\lambda_1$ & $\lambda_2$ & $x_r$ & $\tau$ \\
 \hline
 Fig.~\ref{Fig:Fig2}(a) & 0 & 0 & 0 & 0.97	\\
 Fig.~\ref{Fig:Fig2}(b) & 1.7 & 1.7 & 0.83 & 0.45\\ 
 Fig.~\ref{Fig:Fig2}(c), Fig.~\ref{Fig:FigLongJunction} & 1.7 & 3.56 & 0.83	& 0.45	\\ 
 Fig.~\ref{Fig:Fig7}(c), Fig.~\ref{Fig:FitPRX2} & 1.3 &	2.3 & 0.525	& 0.295	\\
 Fig.~\ref{Fig:FitPRX1} & 2.80 &	4.85	& 0.18	& 0.25 	\\
 Fig.~\ref{Fig:Fig11} & 1.86 &	1.86	& 0	& 0.996	\\
 ~ & 0.95 &	0.95 & 	-1	& 0.26 \\
 \hline
 \end{tabular}
\end{center}
\caption{Parameters for the spectra shown in the figures, calculated using Eq.~(\ref{eq:transcendental}).}
\label{Table:spiderfit}
\end{table}

\acknowledgments 
Technical support from P. S\'enat is gratefully acknowledged. We thank P. Orfila and S. Delprat for nanofabrication support, and our colleagues from the Quantronics group and C. Strunk for useful discussions. We thank M. Hays and V. Fatemi for their in-depth reading of our manuscript and for sharing with us Ref.~\cite{Hays2020}. 
This work has been supported by ANR contract JETS, by FET-Open contract AndQC, by the Renatech network, by the Spanish AEI through Grant No.~FIS2017-84860-R 
and through the ``Mar\'{\i}a de Maeztu'' Programme for Units of Excellence in R\&D (CEX2018-000805-M).
L. Tosi was supported by the Marie Sk\l{}odowska-Curie individual fellowship grant 705467. C. Metgzer was supported by Region Ile-de-France in the framework of DIM SIRTEQ. A. A. Reynoso acknowledges support from projects No.~FIS2017-86478-P (MINECO, Spain) and E041-01 No.~174-2018-FONDECYT-BM-IADT-AV (CONCYTEC, Perú).

C.M., S.P. and L.T. contributed equally to this work.

\end{document}